\documentclass{egpubl}

\JournalSubmission    
\usepackage[T1]{fontenc}
\usepackage{dfadobe}  

\usepackage{cite}  

\usepackage[utf8]{inputenc}
\usepackage{xcolor}
\usepackage{subcaption}
\usepackage{booktabs} 

\usepackage[mathlines, switch]{lineno}

\usepackage{url}

\usepackage{kotex}
\usepackage{mathtools}
\usepackage{multirow}
\usepackage[ruled]{algorithm2e} 

\usepackage{colortbl}
\usepackage{tabulary}

\BibtexOrBiblatex
\electronicVersion
\PrintedOrElectronic
\ifpdf \usepackage[pdftex]{graphicx} \pdfcompresslevel=9
\else \usepackage[dvips]{graphicx} \fi

\usepackage{egweblnk} 

\title[Understanding the Stability of Deep Control Policies for Biped Locomotion]%
      {Understanding the Stability of Deep Control Policies for Biped Locomotion}

\author[Hwangpil Park et al.]
{\parbox{\textwidth}{
\centering
Hwangpil Park$^{1}$,
Ri Yu$^{1}$,
Yoonsang Lee$^{2}$,
Kyungho Lee$^{3}$
and Jehee Lee\thanks{Corresponding Author, jehee@mrl.snu.ac.kr}$^{1}$
        }
        \\
{\parbox{\textwidth}{
\centering
$^1$Department of Computer Science and Enginering, Seoul National University, Computer South Korea\\
$^2$Department of computer Science, Hanyang University, South Korea\\
$^3$NC Soft, South Korea
      }
}
}

\begin{document}

\teaser{
\includegraphics[width=.90\textwidth]{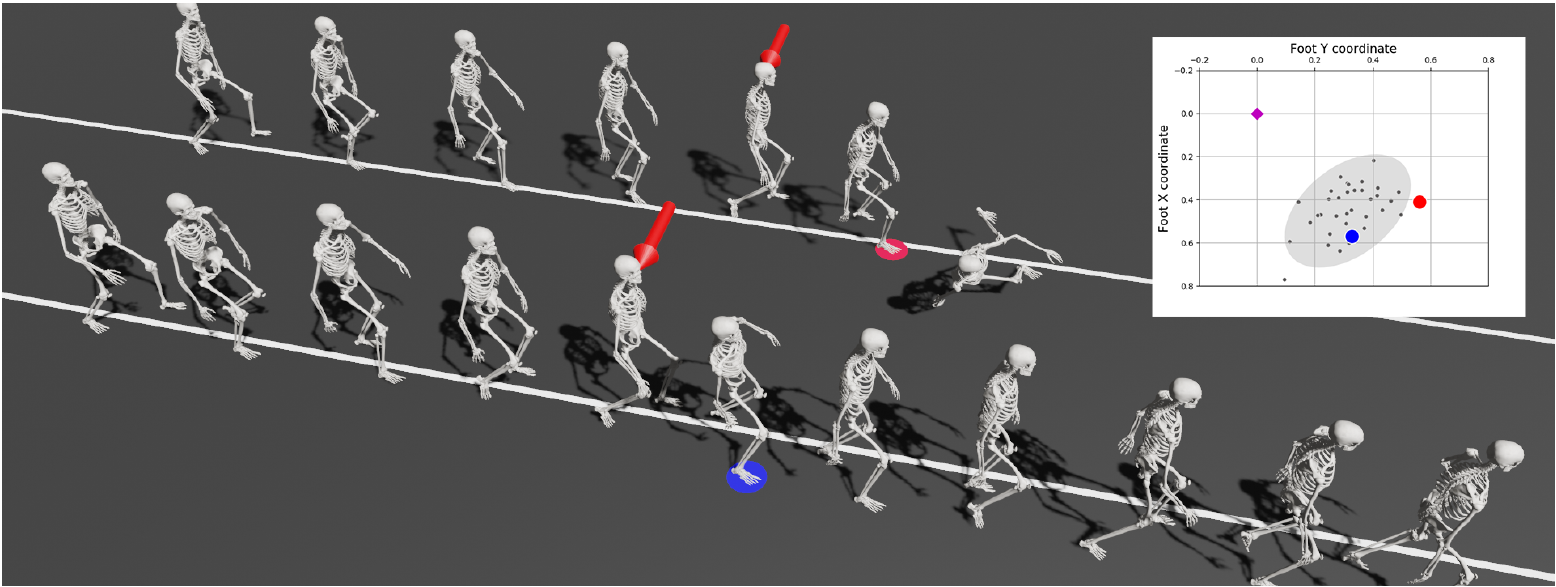}    
\centering
\caption{
Push-recovery experiments. The resilience and robustness of  biped locomotion can be qualitatively characterized and quantitatively assessed from the analysis of the post-push foot placements. Our experiments show that DRL policies are as robust as human walking.
}    
\label{fig:teaser}
}

\maketitle
\begin{abstract}
Achieving stability and robustness is the primary goal of biped locomotion control. Recently, deep reinforce learning (DRL) has attracted great attention as a general methodology for constructing biped control policies and demonstrated significant improvements over the previous state-of-the-art. Although deep control policies have advantages over previous controller design approaches, many questions remain unanswered. Are deep control policies as robust as human walking? Does simulated walking use similar strategies as human walking to maintain balance? Does a particular gait pattern similarly affect human and simulated walking? What do deep policies learn to achieve improved gait stability? The goal of this study is to answer these questions by evaluating the push-recovery stability of deep policies compared to human subjects and a previous feedback controller. We also conducted experiments to evaluate the effectiveness of variants of DRL algorithms.

%
%

\end{abstract}  
\section{Introduction}

The simulation and control of human locomotion is a central issue in physically-based animation. The design principles of biped controllers have pursued several fundamental goals: Simulated locomotion should look human-like, should be resilient against unexpected disturbances, and should be interactively controllable to change its walking direction and speed. It is also desirable for the controller to be able to generate various gait patterns and transitions. We are particularly interested in the second goal and understanding the balance recovering capability of biped controllers. 

Many design approaches have been explored to construct robust biped controllers: feedback control laws~\cite{conf/siggraph/HodginsWBO95, yin2007simbicon}, data-driven control (in the sense that the controller mimics a reference motion)~\cite{sok2007simulating, lee2010data, 2016-TOG-controlGraphs}, nonlinear/stochastic optimization~\cite{Lasa:2010:SIGGRAPH, Wang:2010:SIGGRAPH, kwon_momentum-mapped_2017}, model-based optimization~\cite{Coros:2010:GBW:1778765.1781156, Mordatch:2010:SIGGRAPH, tsai2009real}, and reinforcement learning~\cite{2018-TOG-deepMimic, yu2018learning}. In particular, recent advances in deep reinforcement learning (DRL) have made significant improvements in the simulation and control of biped locomotion. There are many variants of DRL algorithms for learning locomotion control. A typical example-guided algorithm takes a short motion clip as input and learns a control policy (a.k.a. controller) that allows the biped to imitate the dynamic behaviors captured in the motion data~\cite{2018-TOG-deepMimic}. Alternatively, DRL algorithms can produce control policies conditioned by continuous, user-controllable gait parameters, which may include walking speeds, steering angles, body shape variations (e.g., leg/arm lengths), and a large repertoire of action choices captured in unorganized motion data sets~\cite{Won:2019, bergamin2019drecon, Park:2019}.  

Although deep control policies are substantially more robust than the previous state-of-the-art, many questions remain unanswered. Are deep control policies as robust as the balance-recovering capability of human locomotion? Do simulated locomotion use similar strategies as human locomotion to maintain balance? Does a particular gait pattern similarly affect human and simulated locomotion? Are conditioned control policies as robust as unconditioned control policies without adjustable parameters? What do deep policies learn to achieve robustness in locomotion? 


In this paper, we evaluate the push-recovery stability of deep policies under various conditions (e.g., walking speed, stride length, the level of crouch, push timing, and push force). To do so, we conducted simulation-based stability tests with each control policy we trained and compared its stability with previous human/simulation experiments. The push-recovery stability measures how well the simulated biped withstands impulsive pushes. More specifically, two measures are adopted in our experiments: maximum detour distance and fall-over rate. 
The detour distance measures how far the biped detours in the direction of modest pushes to assess the resilience of human walking without falling over. The fall-over rate is more popular for assessing the stability of simulated controllers because the experiments can be conducted safely in the simulation environment with a wider range of push magnitudes. The characteristics of control policies are analyzed qualitatively and quantitatively based on post-push foot placement patterns. 

We also evaluate the effectiveness of DRL variants. Gait-conditioned policies have many advantages over unconditioned, gait-specific policies in terms of computational time and memory usage. It has been believed that those advantages are gained at the cost of sacrificing the balance capabilities to some extent. In our experiments, we observed that gait-conditioned policies are not necessarily inferior to gait-specific policies in terms of push-recovery stability. It has been also found that adaptive sampling in the gait parameter domain results in more robust policies than na{\"i}ve non-adaptive learning, and learning with random pushes results in more robust policies than learning without random pushes. Random disturbances in the learning process not only improve resilience but also allow DRL policies to better emulate human balance strategies in the foot placement analysis. Overall, we found that DRL policies are as robust as human walking.

\section{Related work}

\subsection{Physics-based Simulation and Control}

The design of biped controllers that produce realistic human walking has been a challenging subject in computer graphics.
The key challenge is designing a balancing mechanism, which is usually implemented as a feedback loop that adjusts the controller output (joint torques or PD target poses) based on its input (body state and environment information). A variety of control approaches has been explored to generate responsive and realistic human locomotion with diverse feedback mechanisms. Finite state machines equipped with manually-designed, intuitive feedback rules have been an effective approach in the early biped controller design~\cite{conf/siggraph/HodginsWBO95,yin2007simbicon}. 
To mitigate the complexity of full-body dynamics, simplified dynamics models such as inverted pendulums have been used in controller design~\cite{kwon2010control, Coros:2010:GBW:1778765.1781156,tsai2009real,Mordatch:2010:SIGGRAPH,kwon_momentum-mapped_2017}.

Data-driven (a.k.a. example-guided) approaches have been frequently used to improve the naturalness of simulated animations by adopting motion capture data as a reference to track~\cite{sok2007simulating,lee2010data,liu2012terrain,2016-TOG-controlGraphs}.
The balance mechanisms for data-driven controllers can be manually-crafted~\cite{lee2010data}, learned from a collection of example motions using a regression method~\cite{sok2007simulating}, or derived from a linear feedback policy using stochastic optimization and/or linear regression~\cite{liu2012terrain,2016-TOG-controlGraphs}.
Model predictive control (MPC) is used for synthesizing the full-body character animations in a physically plausible manner using reference motion data \cite{da2008simulation, hong2019physics}. Trajectory optimization is also employed to fulfill given tasks \cite{al2012trajectory, pan2018active, liu2018learning}. Many studies have utilized nonlinear optimization methods to  improve the robustness of controllers, or to explore control schemes for given tasks ~\cite{ye2010optimal, sok2007simulating,liu2012terrain,Lasa:2010:SIGGRAPH, Wang:2010:SIGGRAPH, wang2012optimizing}.

\subsection{DRL for Locomotion Control}

Recently, deep reinforcement learning has received significant attention and shown impressive improvements in the biped control problem. 
The control policy represented by a deep neural network can effectively achieve a feedback balancing mechanism in learning-based control.
A variety of DRL algorithms has been proposed to learn control policies for biped locomotion~\cite{heess_emergence_2017, schulman2015high, yu2018learning}. Yu et al.~\shortcite{yu2018learning} proposed a mirror symmetry loss with curriculum learning to learn a locomotion policy without any reference data.

DRL algorithms can also take advantage of using motion capture data. Liu et al.~\shortcite{liu_learning_2017} used deep Q-learning to learn a scheduler that reorders short control fragments which are in charge of reproducing short motion segments. Peng et al.~\shortcite{peng_deeploco:_2017} presented a two-level hierarchical DRL-based control framework learned from short reference motion clips. They also presented a DRL method in their follow-up study to learn a control policy that imitates a given reference motion clip~\cite{2018-TOG-deepMimic}.
Lee et al.~\shortcite{lee2019scalable} proposed a DRL-based controller for a muscle-actuated anatomical model.
Learning a control policy that exploits a set of reference motion data can be facilitated by recurrent neural networks~\cite{Park:2019} and a motion matching technique~\cite{bergamin2019drecon}.
Although the adoption of DRL in biped control has been very successful, few in-depth analysis of the characteristics of DRL-based controllers has been presented.

\subsection{Stability Analysis}

Gait and postural stability has been measured quantitatively using a waist-pull system~\cite{rogers2001lateral} and a movable platform~\cite{brauer2001interacting}, which can apply quantified perturbations to human subjects. In biomechanics, gait stability has been estimated using measures derived from nonlinear time-series analysis, such as Lyapunov exponents and Floquet multipliers~\cite{dingwell2000local}. The correlation of foot placement with balancing capability has been investigated in biomechanics and robotics~\cite{10.1115/1.4005462}. Wight et al.~\shortcite{10.1115/1.2815334} advocated Foot Placement Estimator (FPE) as a measure of balance for bipedal robots. 
In computer graphics, measuring the response to unexpected external pushes is a common criterion for estimating the resilience and robustness of controllers~\cite{yin2007simbicon,Mordatch:2010:SIGGRAPH,kwon_momentum-mapped_2017,lee2010data,2016-TOG-controlGraphs}.



Lee et al.~\shortcite{lee2015push} statistically analyzed the push-recovery stability of humans and that of simulated bipeds controlled by a hand-crafted feedback controller~\shortcite{lee2010data}.
Using maximum detour distance as a stability measure, they identified key gait factors (walking speed, level of crouching, push magnitude and timing) that affected the stability of human walking through statistical analysis. Their experimental setups are adopted in our simulation experiments and their human experiments serve as a reference of human vs simulation comparisons in our study.



This study begins with a question about how the characteristics of DRL-based controllers are similar or different compared to human walking and previous biped controllers. We aim to gain a deeper understanding and insight into how DRL achieves better stability in notoriously-challenging control problems.

\section{Biped Locomotion Simulation}

\begin{figure}
    \centering
    \includegraphics[width=\linewidth]{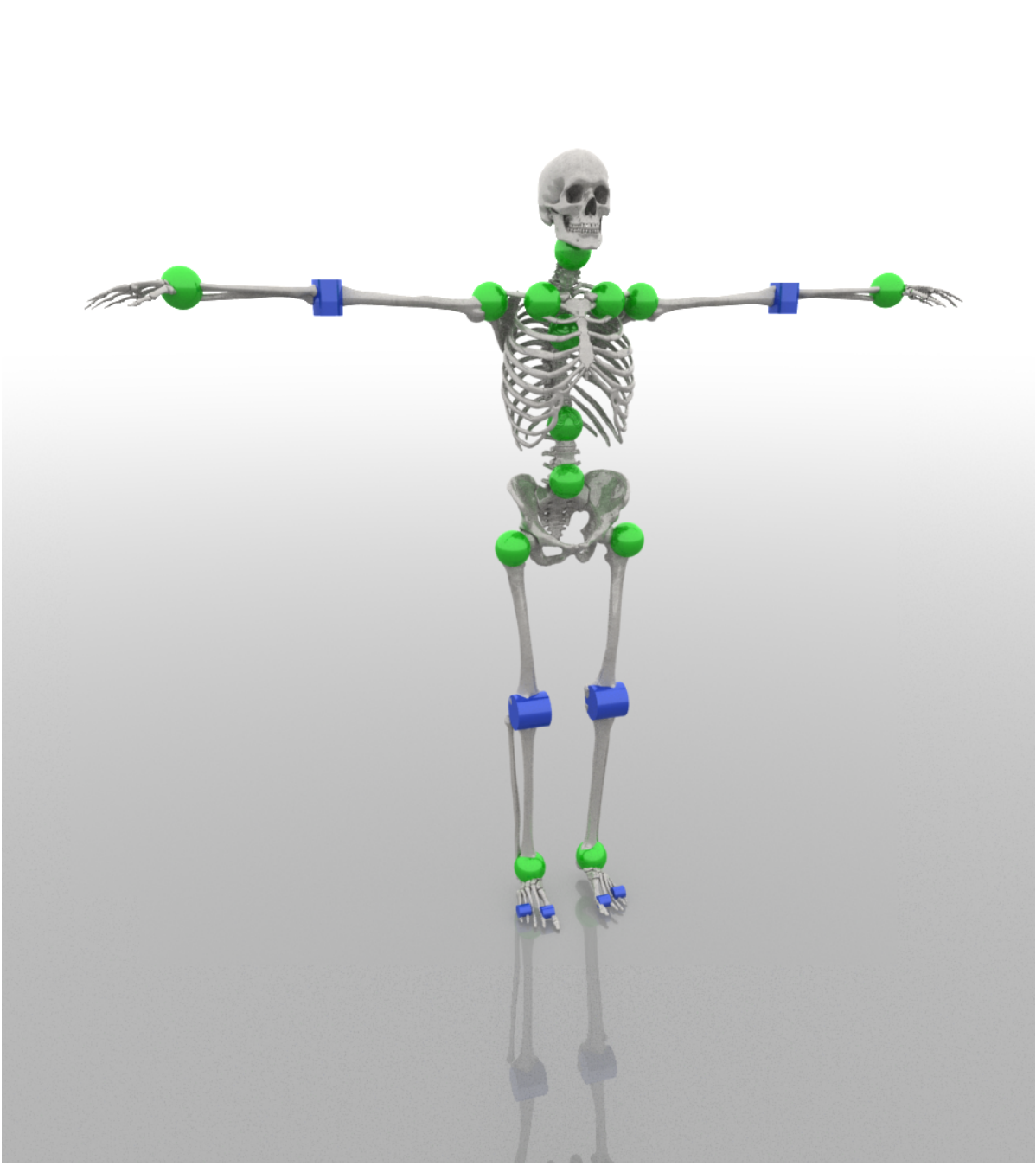}
    \caption{Our full-body dynamics model. Green balls and blue cylinders represent ball-and-socket (3 DoF) joints and revolute (1 DoF) joints, respectively.}
    \label{fig:model}
\end{figure}

Our full-body biped model has rigid bones connected by 8 revolute joints (elbows, knees, and toes) and 14 ball-and-socket joints. The total degrees of freedom (DoFs) of the model is 56 with unactuated 6-DoFs root joint. The biped is 1.7 meters tall and weighs 72 kg. The articulated skeleton is physically simulated and actuated by joint torques (see Figure~\ref{fig:model}).

In this study, we consider two versions of DRL-based algorithms for stability analysis: gait-specific and gait-conditioned. The gait-specific algorithm serves as a common component of recent biped simulation studies \cite{2018-TOG-deepMimic,lee2019scalable}. It takes a short reference motion clip as input and learns a control policy using reinforcement learning. This example-guided control policy represented by a deep neural network provides a distribution of plausible actions at every state. A series of plausible actions sampled from the distributions would drive the biped to track the reference motion while maintaining its balance.

The gait-conditioned algorithm exploits a family of reference motion clips parameterized by user-controlled parameters, such as walking speed and stride length. The gait-conditioned control policy learns how to deal with variations in gaits and styles. Learning a gait-conditioned policy is more computation-efficient and memory-efficient than learning a grid of gait-specific policies in the parametric domain, since each individual gait-specific policy should be learned from scratch. 

\subsection{Gait-specific Policies}

The state of the biped $s=[p,\dot p,\phi]$ includes the positions and velocities of the body links relative to the body coordinate system attached to the skeletal root. $\phi\in[0,1]$ is a normalized phase that matches the temporal span of the reference motion. The output of the control policy is a PD (Proportional Derivative) target that generates joint torques through PD control. The reward function is
\begin{equation}
    r = ( {w}_{q}{r}_{q} + {w}_{v}{r}_{v} ){r}_{e} + {w}_{g}{r}_{g},    
\end{equation}
where ${r}_{q}$, ${r}_{v}$, and ${r}_{e}$ are the reward for tracking reference poses, joint velocities, and positions of end-effectors, respectively. The end-effector and position/velocity tracking rewards are multiplied as suggested by Lee et al.~\shortcite{lee2019scalable}, since they are reinforcing each other. ${r}_{g}$ encourages the biped to walk along a straight line and thus come back to the line after an external push. We set ${w}_{q} = 0.8$, ${w}_{v} = 0.1$, and ${w}_{g} = 0.1$.
\begin{equation}
\begin{split}
    r_q &= \exp \left( -\frac{\lVert \hat{\mathbf{q}} - \mathbf{q} \rVert ^2}{\sigma_{q}} \right) \\
    r_v &= \exp \left( -\frac{\lVert \hat{\dot{\mathbf{q}}} - \dot{\mathbf{q}} \rVert ^2}{\sigma_{v}} \right) \\
    r_e &= \exp \left( -\frac{\lVert \hat{\mathbf{p}}_{e} - \mathbf{p}_{e} \rVert ^2}{\sigma_{e}} \right) \\
    r_g &= \exp \left( -\frac{\lVert \mathrm{dist}(\mathbf{d}, \mathbf{p}_{c}) \rVert ^2}{\sigma_{g}} \right).
\end{split}
\end{equation}
Here, $\mathbf{q}$ is an aggregated joint angle vector, $\dot{\mathbf{q}}$ is a time derivative of $\mathbf{q}$, and $\mathbf{p}_{e}$ is an aggregated vector of end-effector positions.
The hat symbol indicates values from the reference motion.  $\mathrm{dist}(\mathbf{d}, \mathbf{p}_{c})$ is the distance from the center of mass (CoM) of the character $\mathbf{p}_{c}$ to the straight line $\mathbf{d}$.

We used Proximal Policy Optimization \cite{schulman2017proximal} and generalized advantage estimation \cite{schulman2015high} to learn a policy function that maximizes the expected cumulative reward. The learning process is episodic. Many experience tuples are collected stochastically from episodic simulations. In each episode, experience tuples are generated by sampling actions from the policy at every time step and the policy is updated systematically by a batch of experience tuples. We refer the readers to the work of Peng et al.~\shortcite{2018-TOG-deepMimic} for implementation details.

Even if the control policy is learned with a reference motion clip, the exploration strategy of reinforcement learning examines unseen states around the reference trajectory to achieve a certain level of resilience to withstand small unexpected disturbances. The control policy can be even more robust if it learns how to cope with disturbances and uncertainty in the learning process. It has been reported that randomly pushing the biped in the episodic simulations would result in improved robustness \cite{Wang:2010:SIGGRAPH, Park:2019}.
In each episode, we applied random force for 0.2 seconds to push the biped sideways from left or right. The detailed values of push magnitude and timing used in the learning process will be described in section 5.



\subsection{Gait-conditioned Policies}

\begin{figure}
    \centering
    \includegraphics[width=.8\linewidth]{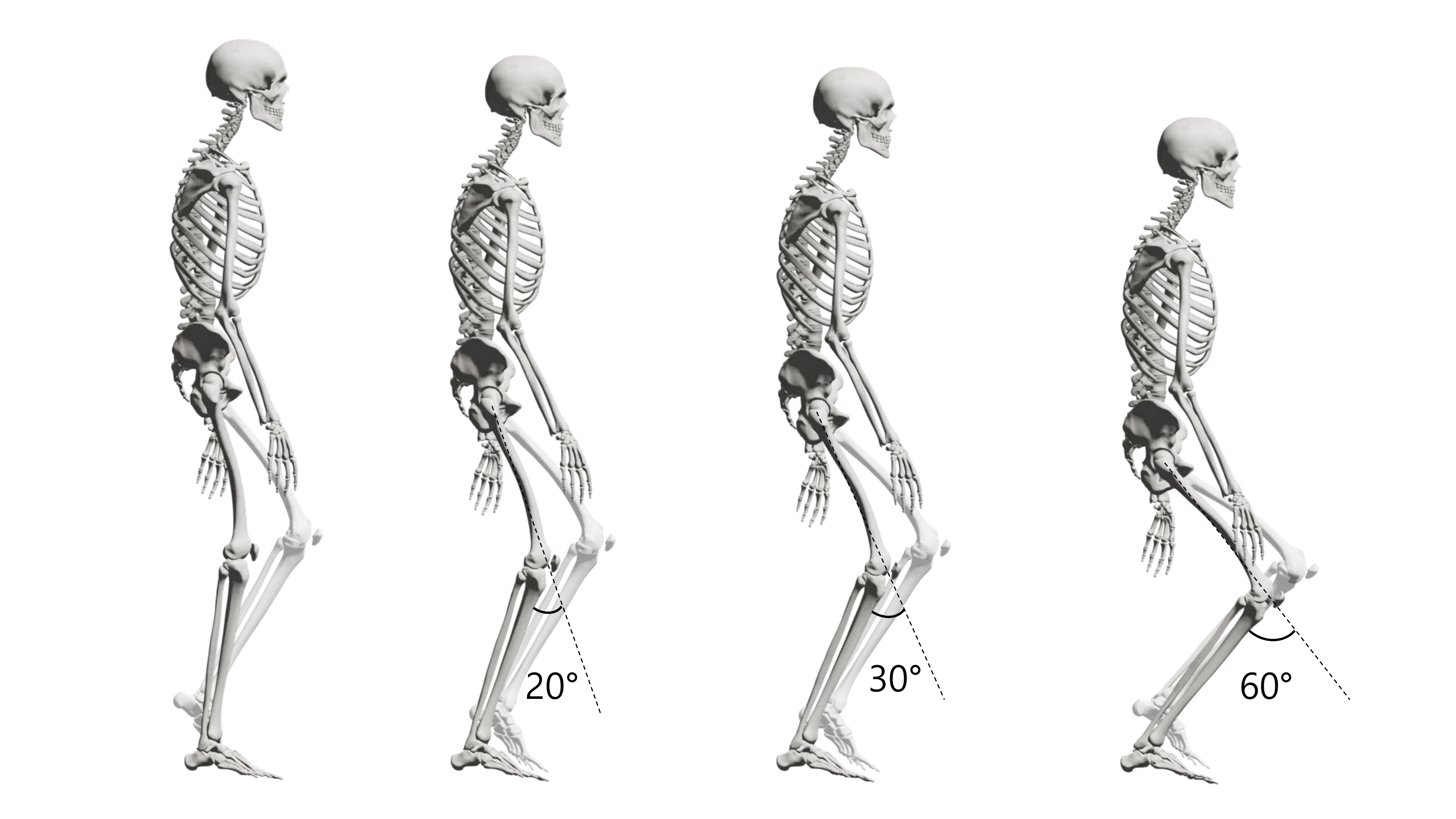}
    \caption{Crouch gaits. The knee angles of normal, $20^{\circ}$, $30^{\circ}$, $60^{\circ}$ crouching at the middle of the right foot stance phase. }
    \label{fig:crouch_level}
\end{figure}

Human locomotion can be characterized by a family of parameters. The flexibility and representation power of deep neural networks allow parametric variations in gaits to be learned in a single network-based policy, which takes those parameters as state input. The state of the gait-conditioned policy $s=[p,\dot p,\phi,c,l,v]$ includes three gait parameters $c,l,v$, which are normalized crouch angle, stride length, and walking speed, respectively. The stance knee is supposed to be straight at the middle of the stance phase in normal gait. The crouch gait has its knee flexed throughout the stance phase (see Figure~\ref{fig:crouch_level}). The crouch angle indicates the level of crouching normalized to $[0,1]$. The stride length and walking speed are normalized to have zero mean and one standard deviation. 

Because a gait-conditioned policy should deal with gait parameters as state input, learning a gait-conditioned policy requires a collection of example motions that span a target parametric domain. We generate example motions by kinematically varying a single reference motion clip, which represents a normal gait with average walking speed and stride length. Given parameter values $(c, l, v)$, the use of hierarchical displacement mapping and time warping edits the reference motion clip to have the desired stride length $l$, crouch angle $c$, and walking speed $v$~\cite{lee1999hierarchical}.

During episodic learning, the initial state of each episode is taken from the state space containing the target parametric domain. A common way to determine the initial state is to sample uniformly in the state space. This na{\"i}ve learning of a policy often suffers from a biased exploration problem. Many DRL algorithms tend to explore successful regions in the target domain more aggressively while leaving less successful regions unexplored in starvation. For example, human walking is more stable when it crouches to lower down its CoM~\cite{Schafer87}. The control policy parameterized by a crouch angle would explore crouch walking more frequently than normal (straight-knee-at-stance) walking. Consequently, the learned normal gait in the policy would be less robust.

Recently, Won et al.~\shortcite{Won:2019} proposed an adaptive sampling method to deal with parametric body shape variations. We adopt their sampling idea to learn our gait-conditioned policies. The key idea is to give more opportunities to less successful regions in the target domain. The measure of success is a marginal value function $V_m(s_\alpha)$ that estimates the sum of expected rewards for each gait parameter $s_\alpha=(c,l,v)$.
\begin{equation}
    V_m (s_{\alpha})= \int_{{S}_{\beta}} V(s_{\alpha}, s_{\beta}) p_{s}(s_{\alpha}, s_{\beta}) d s_{\beta},
\end{equation}
where $V(s)=V(s_\alpha,s_\beta)$ is a value function which approximately measures the cumulative reward when the controller follows current policy from state $s$, $s_{\beta}$ is a state vector excluding gait parameters, $S_{\beta}$ is the domain of $s_{\beta}$, and $p_{s}$ is a density function which is assumed a constant. The probability of exploring $P ({s}_{\alpha})$ is 
\begin{equation}
    P ({s}_{\alpha}) = \frac{1}{Z}\exp \left( -k \left( \frac{V_{m} ({s}_{\alpha}) }{\mu} - 1\right) \right).
\end{equation}
Here, $\mu$ is the expectation of $V_m$, which is updated along with $V_m$, $Z$ is a scaling factor to normalize $P$, and $k$ is the value that decides the degree of uniformity of $P$ over gait parameter space. MCMC (Markov Chain Monte Carlo) sampling with the probability aims to make the marginal value function $V_m(s_\alpha)$ near-uniform across the domain of $s_\alpha$. We chose $k = 1$ for all controllers.

\section{Push-Recovery Experiments}

The primary goal of this research is to assess the robustness of deep policies in comparison with human subjects and pre-deep learning controllers. The study of Lee et al.~\shortcite{lee2015push} provides a reference for the comparison. We conduct push-recovery experiments with identical simulation setups to generate measurement data that can be directly compared to their results.

\begin{figure} 
  \includegraphics[width=\linewidth]{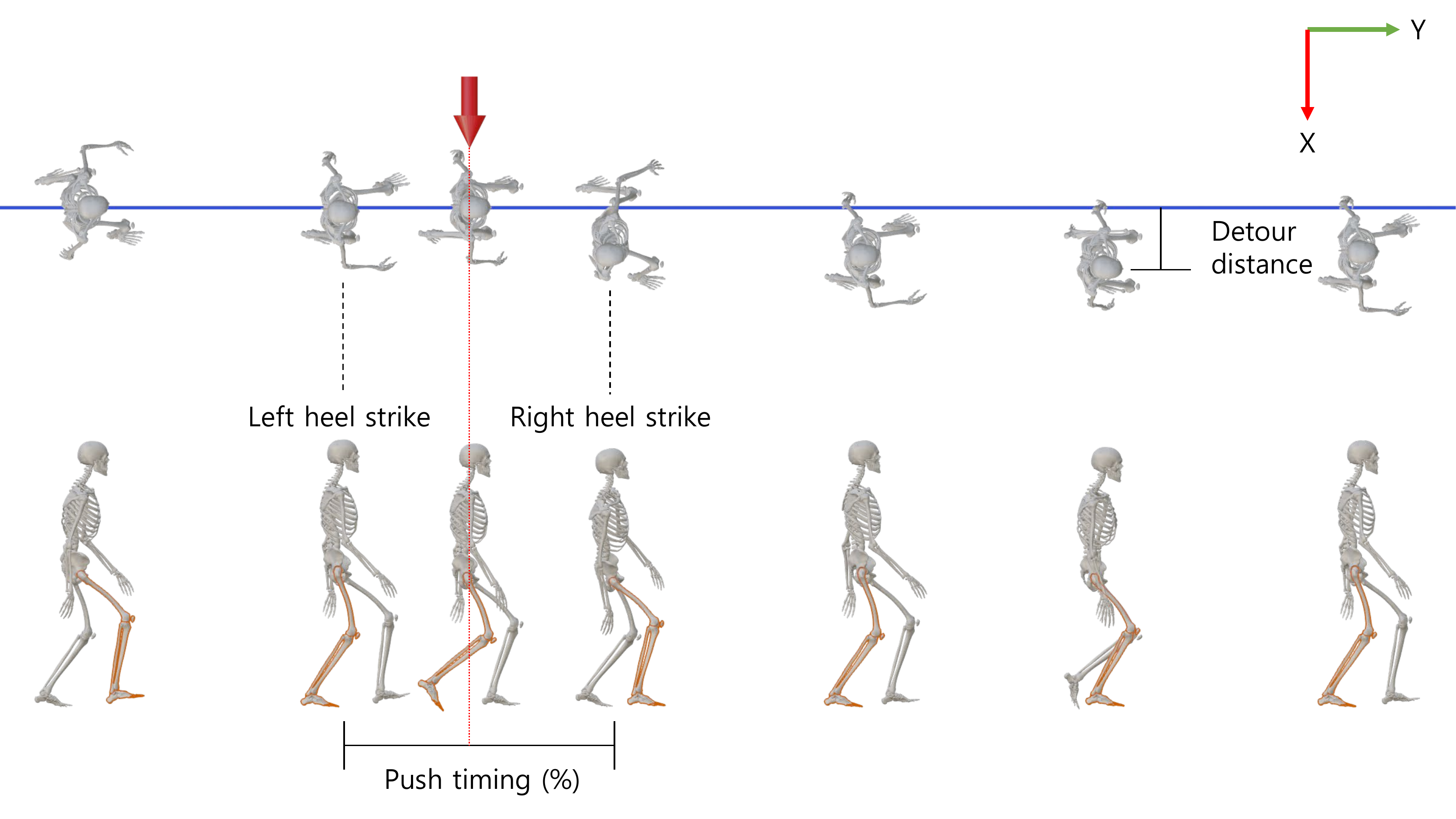}
  \caption{The setup for push-recovery experiments from top and lateral views. The blue line indicates the heading direction (Y axis).}
  \label{fig:push_recovery}
\end{figure}

The motion capture and measurement data from the previous study are available on their project webpage ~\cite{lee2015push}. The data set includes push-recovery experiments of 29 healthy adults (14 males and 15 females). The participants walked along a straight line with the choice of two speeds (normal/slow), two stride lengths (normal/short), and four levels of crouching (normal/20$^\circ$/30$^\circ$/60$^\circ$). The experimenter pushed the participants sideways to measure the maximal detour distance (see Figure~\ref{fig:push_recovery}). They also recorded a great deal of data including the height, weight, leg length, BMI of the participants, magnitude/timing/direction/duration of pushes, number of steps to maximal detour, detour of the first step, and push force normalized by height/weight/leg length. We refer the readers to the work of Lee et al.~\shortcite{lee2015push} for the details of the data acquisition process and specification.

Lee et al.~\shortcite{lee2015push} also performed statistical analysis on their data using a LMM (Linear Mixed Model) method and identified four significant factors (the level of crouch, walking speed, push timing and magnitude) that are correlated with maximal detour distance. Based on the analysis, they also performed comparisons between human and simulation experiments to see if simulated controllers are as robust as human balance strategies. Specifically, they adopted a Data-Driven Controller (DDC) by Lee et al.~\shortcite{lee2010data} for their experiments, which was designed before the advent of deep learning. Their experiments showed that the response pattern of the controller is qualitatively similar to how humans respond to external pushes, though the controller is not as robust as human walking yet.

\begin{figure}
    \centering
    \includegraphics[width=.8\columnwidth]{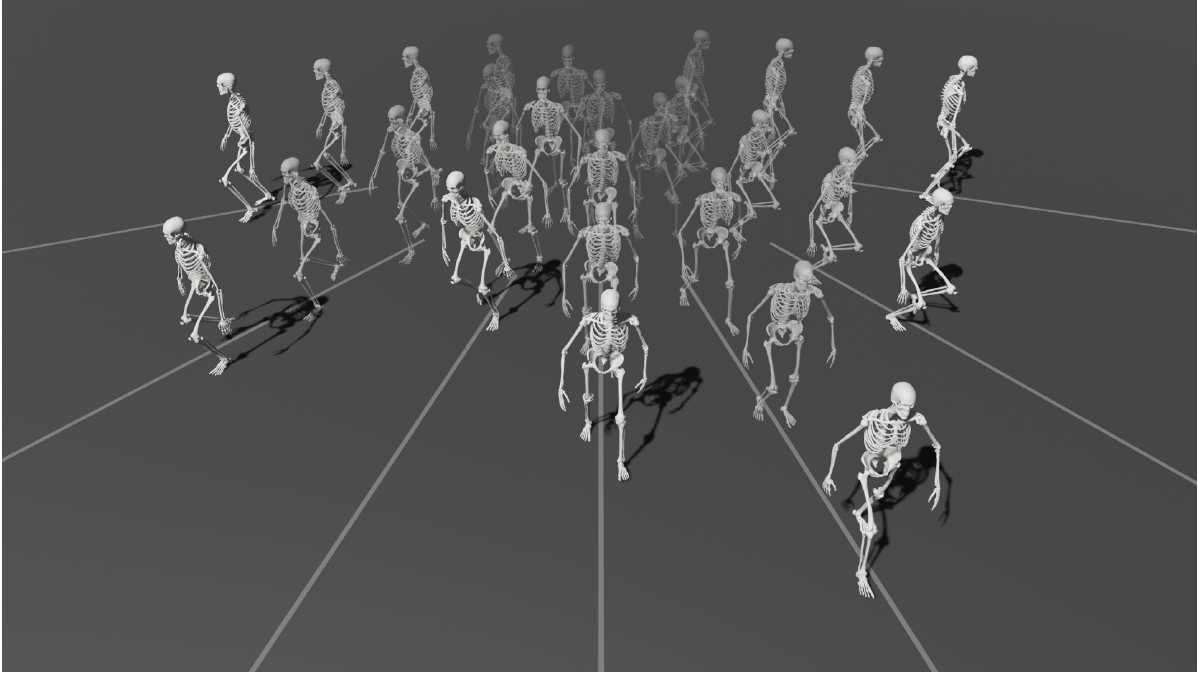}
    \caption{A family of walking motions with various walking speeds, stride lengths, and crouch angles.}
    \label{fig:param_all}
\end{figure}

In this paper, we conduct two sets of experiments. The first type of experiments is to faithfully reproduce the experiments of Lee et al.~\shortcite{lee2015push} with a new state-of-the-art biped locomotion simulation. Through the experiment, we would like to understand how deep policies compare with human walking and pre-deep-learning simulators. To do so, we generated a family of kinematic walking motions with stride lengths, walking speeds, crouch angles, and the magnitude/timing of pushes that match the distribution of human data (see Figure~\ref{fig:param_all}). The crouch angle is discrete, while all the other parameters are continuous and sampled from normal distributions. The push force in the simulation environment is applied to the shoulder and its magnitude is also sampled from the normal distribution of the human experiment data. The push timing is sampled between the left heel strike ($0\%$) and the subsequent right heel strike ($100\%$). Ten thousand push experiments were performed for each of four crouch levels. 

The second type of experiments is for assessing the implementation choices in DRL algorithms, such as random perturbation in learning, gait-specific versus gait-conditioned policies, and adaptive sampling. Since only DRL algorithms are compared with each other, we measure the success rate for evaluating the stability of control policies. An episode of simulation is considered successful if the biped withstands a push and keeps walking for 10 seconds afterwards while maintaining its balance. 

There are several popular methods, such as maximum Lyapunov exponents and maximum Floquet multipliers~\cite{dingwell2000local}, for estimating the stability of dynamical systems. These methods quantify how the dynamical systems respond to small perturbations assuming strict periodicity. Small exponents or multipliers indicate that the system would return to a limit cycle. Human locomotion under disturbances is not a simple dynamical system since the control system is continuously modulated to maintain its balance. Its analysis often requires radical approximation of the dynamical system. In computer graphics, push-recovery stability is more popular since resilience against unexpected disturbances is closer to real-life notions of stability. Push-recovery stability quantifies how humans respond to impulsive perturbations. More specifically, maximum detour distances are useful if the experiment involves human participants who can cope with only moderate disturbances. The rate of success would be a better criterion under larger disturbances. There is no ultimate measure of dynamic stability. Assorted measures illuminate different aspects of gait stability. It is possible that a particular gait is very stable in one criterion but not in another.

\section{Analysis and Results}

We used Intel® Core™ i9-9900K CPU @ 3.60GHz (8 cores) for training. Training takes about 24 hours for gait-specific policies, and 40 to 50 hours for gait-conditioned policies. The neural network for all DRL controllers consists of three fully connected layers with 256 nodes. The network is updated whenever 8192 experience tuples are collected with 256 batch size. We used R (version 3.6.1) for the statistical analysis. 

In our experiments, we compared the push-recovery capabilities of human participants (Human), a data-driven controller (DDC) with hand-crafted feedback laws~\cite{lee2010data}, DRL gait-specific policies (DRL-specific), and DRL gait-conditioned policies with two continuous parameters (walking speed and stride length) and one discrete parameter (crouch angle). We denote gait-conditioned policies (see Table~\ref{tab:types_of_policies}) learned with and without adaptive sampling by DRL-A and DRL-B, respectively. DRL-specific*, DRL-A*, and DRL-B* are policies learned with unexpected disturbances in the learning phase.

\begin{table}
\caption{The name of each control policy. An asterisk means that the policy is trained with push, and -A means that adaptive sampling method is applied in the training.}
\centering
\begin{tabular}{ccc}
\toprule
                  & With push        & Without push \\
\midrule
Adaptive sampling &   DRL-A*         &   DRL-A      \\
\midrule
Uniform sampling  &   DRL-B*         &   DRL-B      \\
\bottomrule
\end{tabular}

\label{tab:types_of_policies}
\end{table}

\begin{table}
\caption{The means and standard deviations of factors in human experiments. The push magnitude is normalized by weight.}
\begin{tabular}{clcc}
\toprule
    \multicolumn{2}{l}{Experimental factor}       & Mean      & Std   \\ \midrule
Walking   &  Normal                               & 0.994     & 0.263 \\
speed     &  Crouch $20\,^{\circ}$                & 0.808     & 0.210 \\
(m/s)     &  Crouch $30\,^{\circ}$                & 0.788     & 0.221 \\
          &  Crouch $60\,^{\circ}$                & 0.744     & 0.228 \\ \midrule
Stride    &  Normal                               & 1.126     & 0.180 \\
length    &  Crouch $20\,^{\circ}$                & 0.953     & 0.158 \\
(m/s)     &  Crouch $30\,^{\circ}$                & 0.916     & 0.167 \\ 
          &  Crouch $60\,^{\circ}$                & 0.876     & 0.168 \\ \midrule
\multicolumn{2}{l}{Push magnitude (N$\cdot$s/kg)} & 0.535     & 0.096 \\ \midrule
\multicolumn{2}{l}{Push timing (\%)}              & 34.0      & 21.0  \\ 
\bottomrule
\end{tabular}
\bigskip\centering
\label{tab:human_data_mean}
\end{table}

The gait parameter domains for learning gait-conditioned policies are decided from human data (see Table~\ref{tab:human_data_mean}). More precisely, at the beginning of every episode in the learning process, gait parameters are sampled uniformly (for DRL-B and DRL-B*) or adaptively (for DRL-A and DRL-A*) in the region within a certain Mahalanobis distance $t$. We chose the value of $t$ to make the sampling region be 95\% confidence region. Note that the walking speed and stride length are correlated, and the correlation coefficient for each crouch walking ranges from 0.67 to 0.83.
For the training of control policies (DRL-specific*, DRL-A*, and DRL-B*) under the circumstance where external pushes are exploited, the simulated character is pushed for 0.2 seconds from the left or the right randomly.
The push direction always matches the stance leg. So, the push from the left happens when the left leg is in stance and vice versa. The magnitude and timing of pushes are sampled to match the distributions in the human experiment data (Table~\ref{tab:human_data_mean}). The push magnitudes were mostly in the range of 100N to 300N.

\subsection{Human vs Simulation}

\color{red} 


\color{black}

The statistics of gait factors in human experiments are shown in Table~\ref{tab:human_data_mean}. The trials in human data are classified into two groups: {\em Group 1} and {\em Group 2}. The participants in {\em Group 1} trials recovered their balance in a single step, and thus the detour distance peaked in the step. Most of the participants recovered their balance within three balance-correcting steps except for a few outliers. They took more than one step when they experienced mild difficulty. In the outliers, participants panicked by unexpected pushes and failed to return to the line. The {\em Group 2} includes all trials except for the outliers.

\begin{figure}
  \includegraphics[width=\columnwidth]{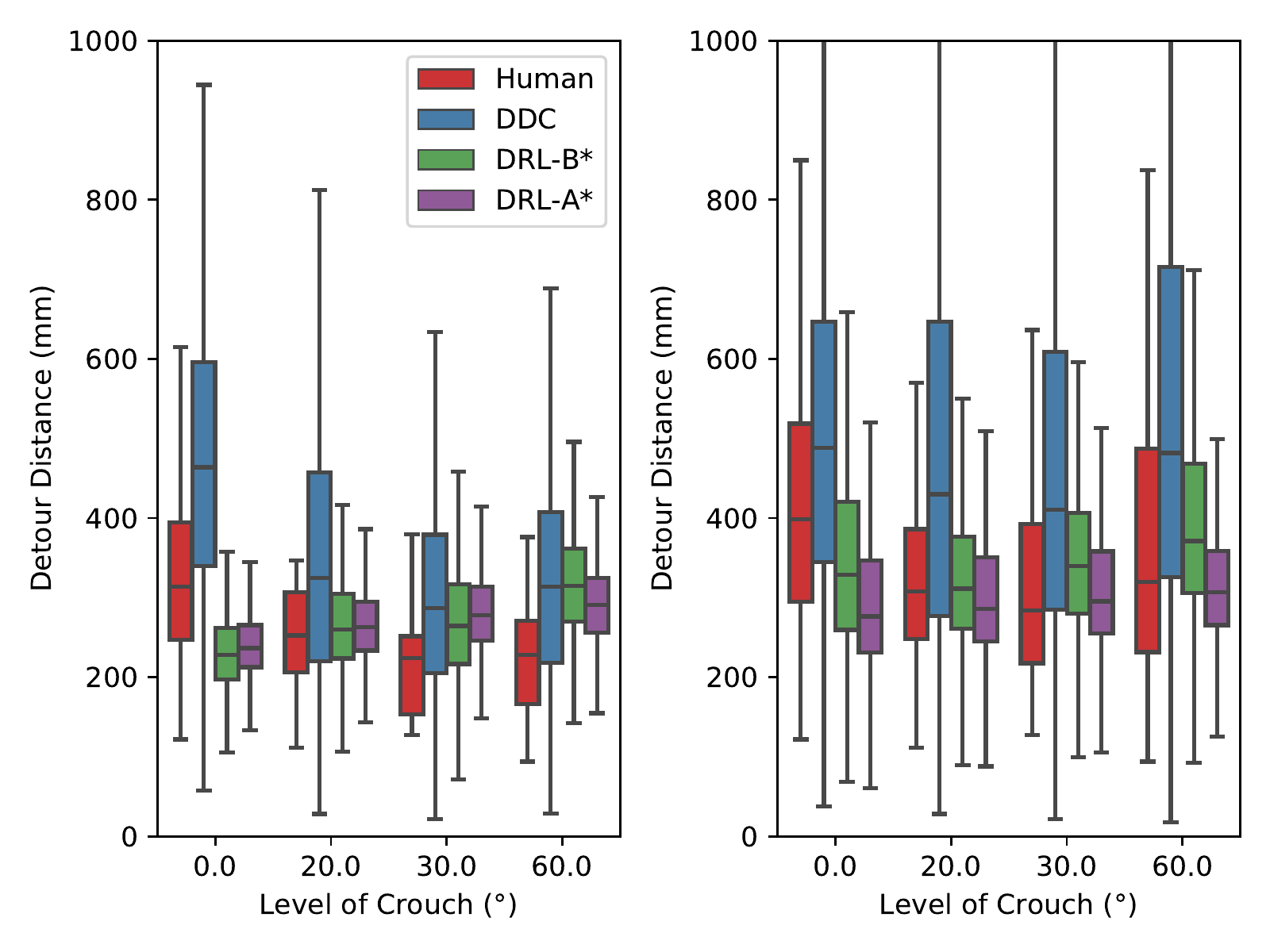}
  \caption{Box plots of maximum detour distance in Group 1 (left) and Group 2 (right). The horizontal line in the middle of each boxplot means median, and colored area represents from 25\% to 75\% of the data.}
  \label{fig:distance_box_plot_each}
\end{figure}

In Figure~\ref{fig:distance_box_plot_each}, we compare the maximum detour distances of Human, DDC, DRL-B*, and DRL-A*. The results of Human and DDC experiments were taken from the work of Lee et al.~\shortcite{lee2015push}. The simulation experiments were designed to reproduce the human experiments with setups and data distributions carefully tuned to match the target experiments. The only difference is that we can collect far more trial data by simulation. 
The {\em Group 1} of Human, DDC, DRL-B*, and DRL-A* include 228, 3858, 4851, and 16036 trials, respectively, and the {\em Group 2} includes 450, 13707, 20534, and 27993 trials, respectively. The comparison graph shows that DRL-B* and DRL-A* are clearly superior to DDC and comparable to human participants.


\begin{figure}
  \includegraphics[width=\columnwidth]{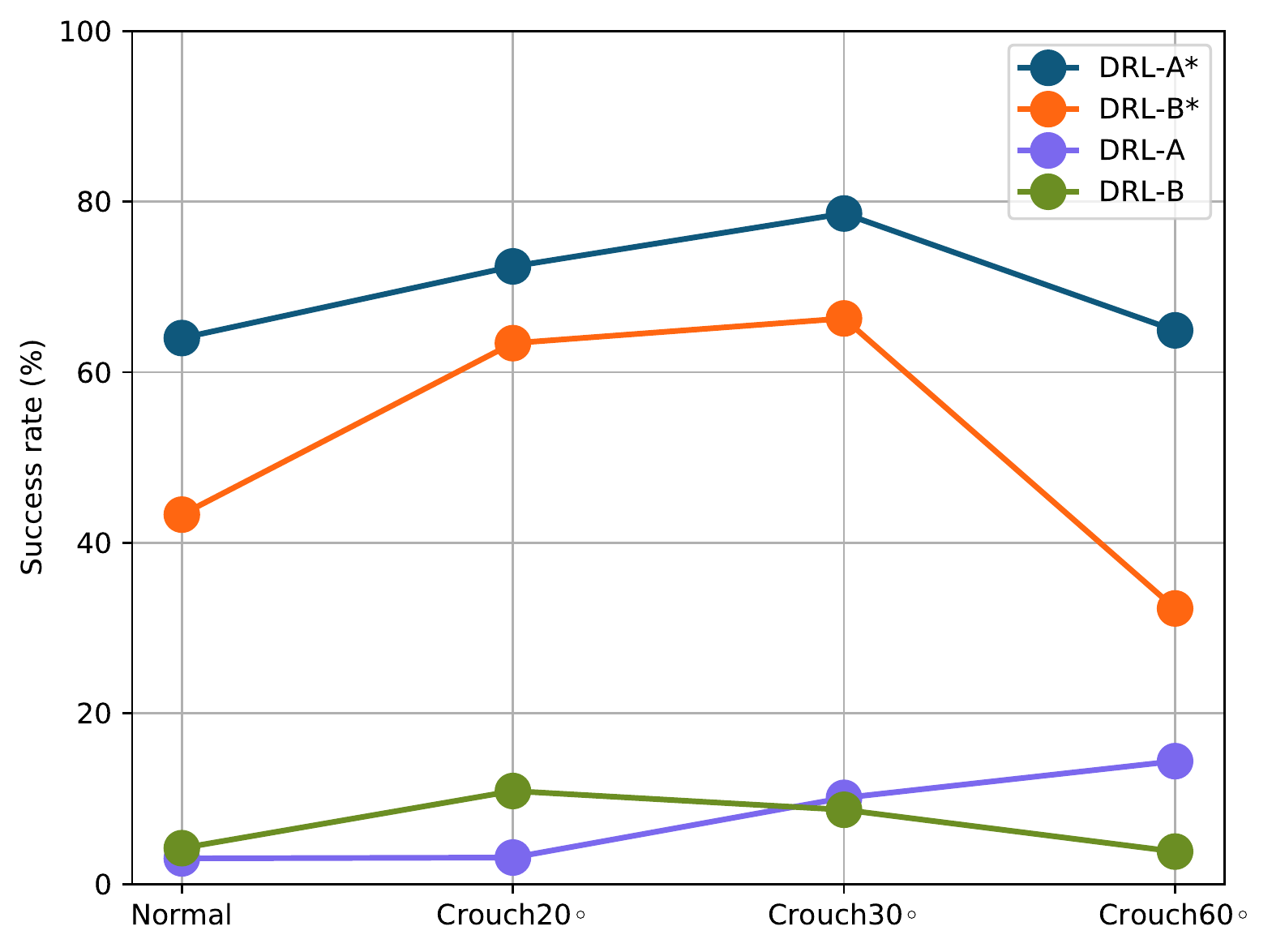}
  \caption{The success rates (\%) of gait-conditioned policies. }
  \label{fig:Sampling_success_rate}
\end{figure}

It was empirically verified in the previous study that crouch walking is more stable than normal walking in Human and DDC experiments. In particular, $30^\circ$-crouch walking was the most stable. This postulation agrees with our intuition that lowering down the CoM improves the gait stability. Although the detour distance measurements of DRL-B* and DRL-A* do not follow this trend, the success rate experiments agree with the postulation. Graph~\ref{fig:Sampling_success_rate} shows that both DRL-B* and DRL-A* are more robust with 20$^\circ$/30$^\circ$-crouch walking than normal and 60$^\circ$-crouch walking. $30^\circ$-crouch walking was the most robust in terms of the success rate.


\begin{table}
\caption{Type 3 tests of the fixed effects (level of crouch, walking speed, magnitude/timing of push) on the detour distance. Significant values ($p < 0.05$) are shown in bold.}
\resizebox{\columnwidth}{!}
{
\begin{tabular}{|c|c|c|c|c|c|c|c|c|c|}
\hline
\multicolumn{2}{|c|}{\multirow{3}{*}{}} & \multicolumn{8}{c|}{Type 3 Tests for fixed effects}                                                                                               \\ \cline{3-10} 
\multicolumn{2}{|c|}{}                  & \multicolumn{2}{c|}{Level of Crouch} & \multicolumn{2}{c|}{Push Magnitude} & \multicolumn{2}{c|}{Walking Speed} & \multicolumn{2}{c|}{Push Timing} \\ \cline{3-10}
\multicolumn{2}{|c|}{}   & $F_{c}$  & \multicolumn{1}{c|}{$p_{c}$}   & $F_{f}$  & \multicolumn{1}{c|}{$p_{f}$} & \multicolumn{1}{c|}{$F_{s}$} & \multicolumn{1}{c|}{$p_{s}$} & \multicolumn{1}{c|}{$F_{t}$} & \multicolumn{1}{c|}{$p_{t}$} \\ \hline\hline
\multirow{4}{*}{Group 1}& Human & \textbf{17.49} & \textbf{<.0001} & \textbf{13.42} & \textbf{0.0003} & 2.68 & 0.1098 & \textbf{14.35} & \textbf{<.0001}   \\ \cline{2-10} 
                    & DDC   & \textbf{30.06} & \textbf{<.0001} & \textbf{17546}  & \textbf{<.0001 } & \textbf{106.6} & \textbf{<.0001} & \textbf{463.4} & \textbf{<.0001}  \\ \cline{2-10} 
                        & DRL-A* & \textbf{196.5} & \textbf{<.0001} & \textbf{11791} & \textbf{<.0001} & \textbf{4536} & \textbf{<.0001} & \textbf{363.5} & \textbf{<.0001}  \\ \cline{2-10} 
                        & DRL-B* & \textbf{768.7} & \textbf{<.0001} & \textbf{3022} & \textbf{<.0001} & \textbf{625.6} & \textbf{<.0001} & \textbf{130.0} & \textbf{<.0001}  \\ 
                        \hline\hline
\multirow{4}{*}{Group 2}& Human & \textbf{8.35} & \textbf{<.0001} & 0.01 & 0.9297 & 0.03 & 0.8578 & \textbf{3.94} & \textbf{0.0479}  \\ \cline{2-10} 
                        & DDC   & \textbf{88.34} & \textbf{<.0001} & \textbf{19103} & \textbf{<.0001} & \textbf{371} & \textbf{<.0001} & \textbf{225.8} & \textbf{<.0001} \\ \cline{2-10}
                        & DRL-A*& \textbf{7.857} & \textbf{0.0051} & \textbf{12427} & \textbf{<.0001} & \textbf{251.8} & \textbf{<.0001} &\textbf{2329} & \textbf{<.0001} \\ \cline{2-10}
                        & DRL-B*& \textbf{575.6} & \textbf{<.0001} & \textbf{10937} & \textbf{<.0001} & \textbf{57.53} & \textbf{<.0001} &\textbf{331.6} & \textbf{<.0001}            \\ \hline
\end{tabular}
}
\label{table:lmm1}
\end{table}

Table ~\ref{table:lmm1} shows the results of Type 3 tests, which shows the fixed effects of level of crouch, walking speed, magnitude/timing of push on the detour distance. The tests confirm that all factors that were proven to be significant in the human experiments are also significant ($p<0.05$) in the DRL experiments.

\begin{figure}
  \includegraphics[width=\columnwidth]{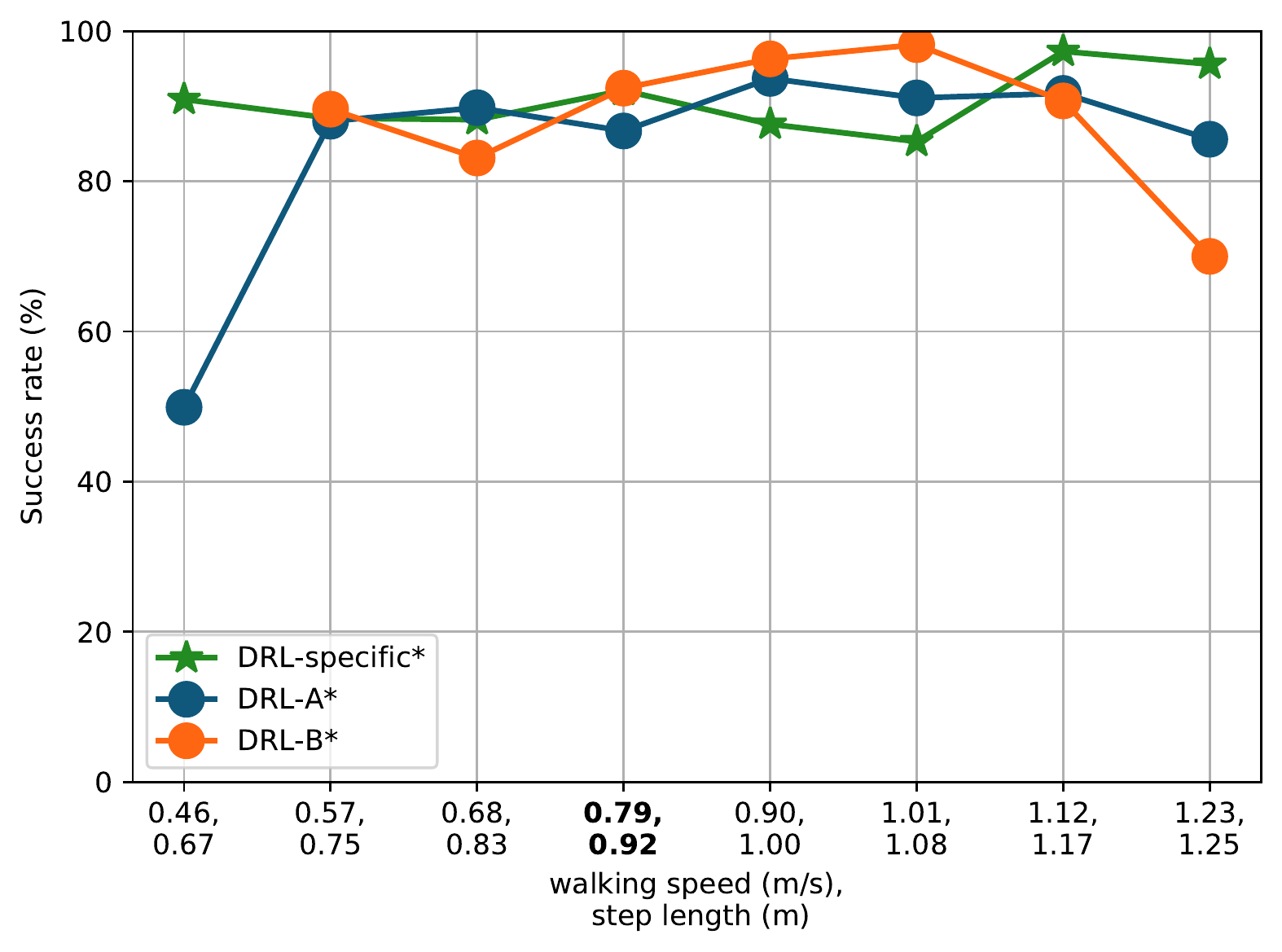}
  \caption{The success rates of gait-specific and gait-conditioned policies of $30^\circ$-crouch walking. Mean parameter values are shown in bold. We did not plot a {\em fail} case where a control policy was unable to learn the walking motion with the corresponding walking speed and stride length.}
  \label{fig:policy_comparison}
\end{figure}

\subsection{Comparison of Control Policies}



Gait-specific policies DRL-specific* supposedly outperform gait-conditioned policies DRL-A* and DRL-B* since the scope of the gait-specific policy focuses on a single reference trajectory, while the gait-conditioned policies have to cope with a continuous spectrum of parametric domains. Specifically, the domain $[\mu_s-1.5\sigma_s,\mu_s+2\sigma_s]\times [\mu_l-1.5\sigma_l,\mu_l+2\sigma_l]$ is explored in our experiments, where $\mu_s$ and $\mu_l$ are the average walking speed and stride length, respectively, in the human experiments. $\sigma_s$ and $\sigma_l$ are their standard derivations. The use of an asymmetric domain is related to the perception plausibility of motion editing. Edited character animations that walk slower than a reference are more likely to look unnatural than animations that walk faster than the reference~\cite{vicovaro_2014}.

\begin{figure}
  \includegraphics[width=\columnwidth]{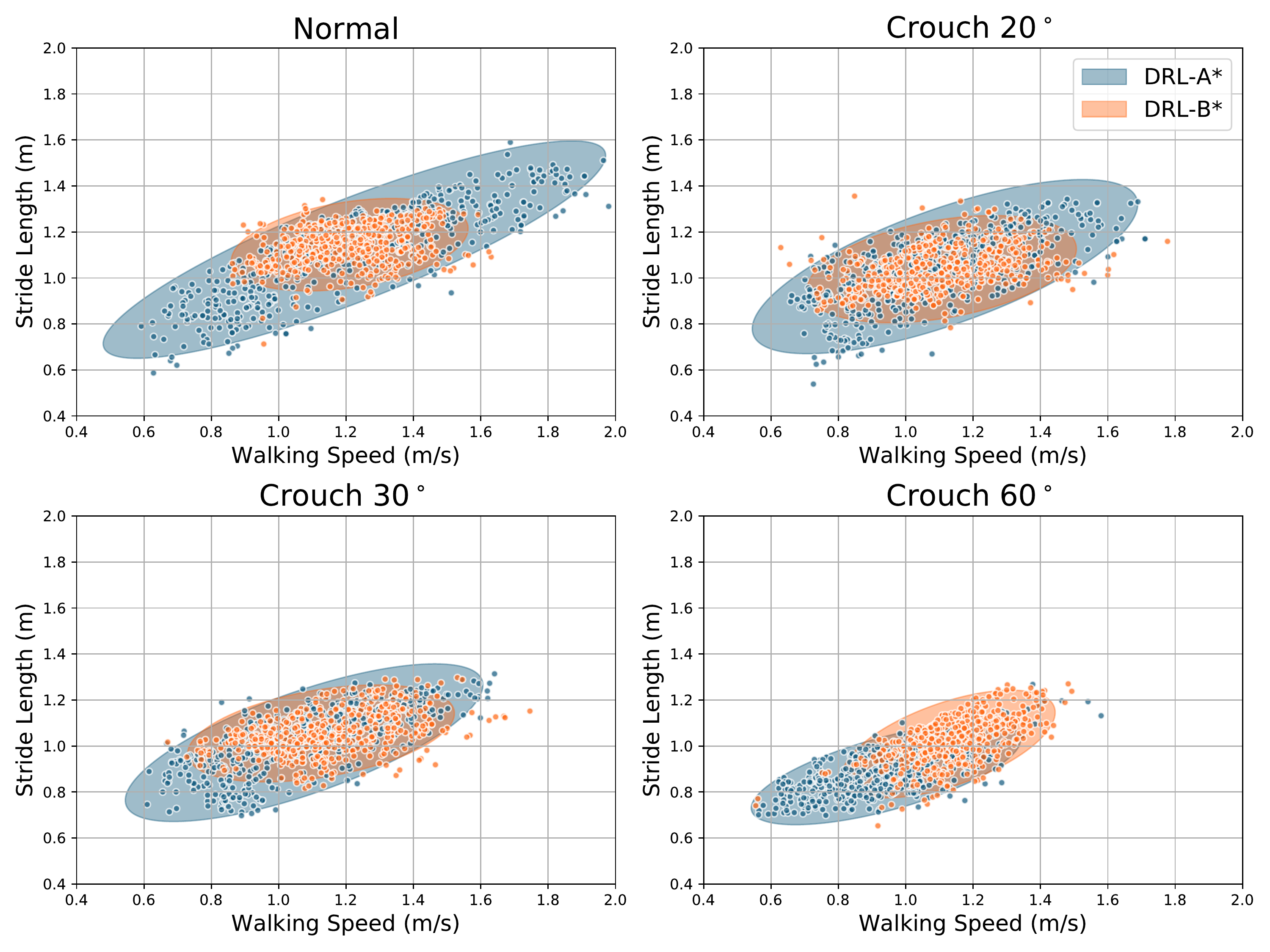}
  \caption{The parameter coverage of successful episodes. Each ellipse represents a 98\% confidence region.}
  \label{fig:sampling_plot}
\end{figure}

We conducted push-recovery experiments of 1000 trials for each of the three policies. Push forces are drawn from a normal distribution with a mean of 200N and a standard derivation of 35N (see Figure~\ref{fig:policy_comparison}). A family of motion clips were generated to learn DRL-specific* polices. The stability of each gait-specific policy learned for a particular walking speed and a stride length is comparable to the stability of DRL-B* near the mean values, but clearly outperforms when the samples are away from the means. It means that na{\"i}ve parametric learning can deal with only a narrow range of the parametric domain. Adaptive sampling of DRL-A* improves the stability at the corners of the domain and consequently learns a policy practically usable over the entire domain within a modest computational time. Figure~\ref{fig:sampling_plot} depicts the parametric coverage of successful episodes. The coverage of DRL-A* is wider than the coverage of DRL-B*. It means that DRL-A* can better deal with slow walking (narrow strides) and fast walking (wide strides), while DRL-B* is effective only at mid ranges. 

Deciding the number of hidden layers of the policy network depends on the dimension and size of the action, state, and parameter spaces. In practice, the number of layers and the number of nodes in each layer are important hyper-parameters to tune empirically. In our experiments, we tested with two, three, and four hidden layers and found that learning was the most successful with three layers.


\subsection{Foot Placement Analysis}

\begin{figure*}
    \begin{subfigure}{.325\textwidth}
      \centering
      \includegraphics[width=\linewidth]{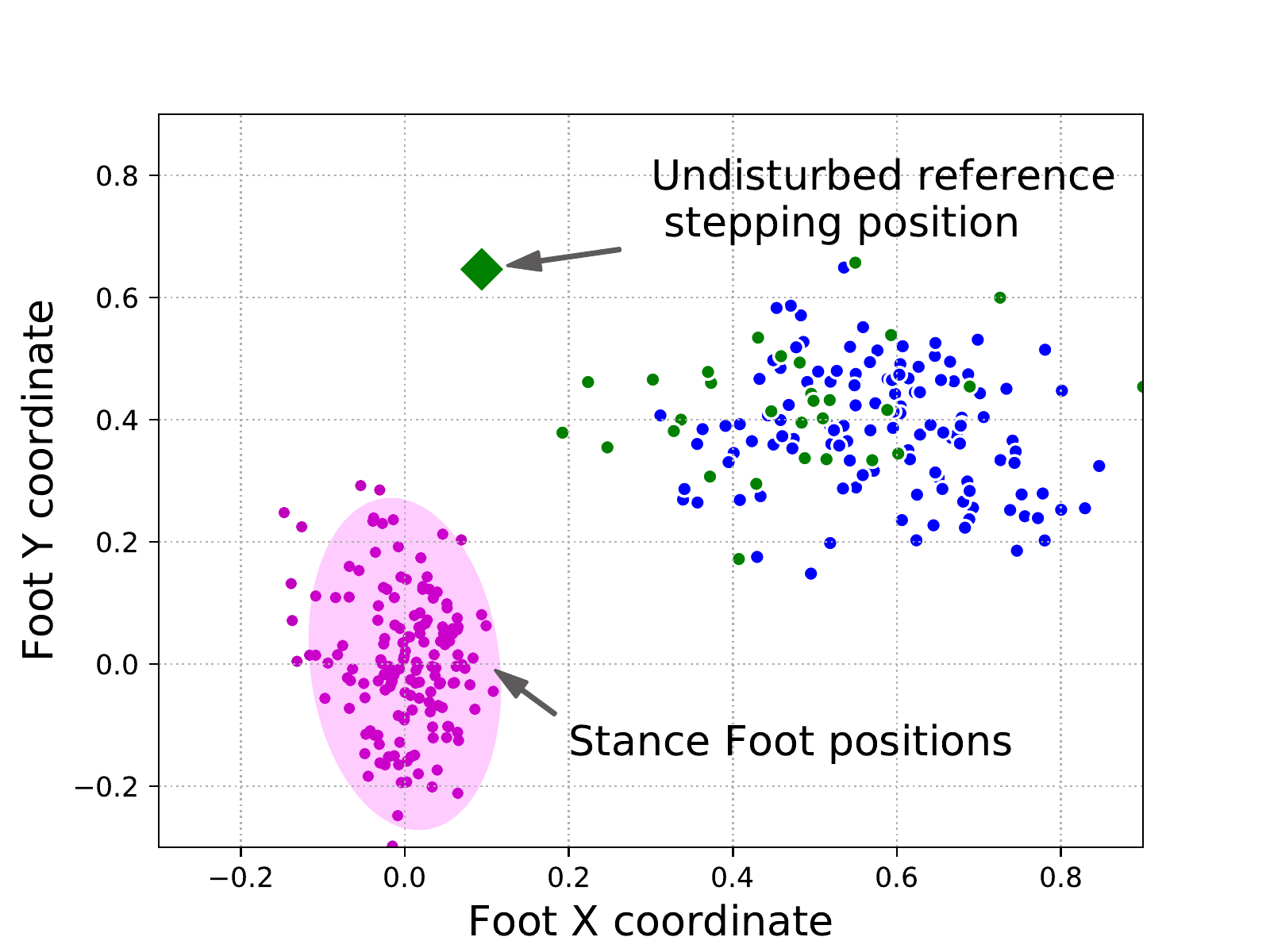}  
      \caption{Human data (Normal)}
      \label{fig:foot_placement_human}
    \end{subfigure}
    \begin{subfigure}{.325\textwidth}
      \centering
    \includegraphics[width=\linewidth]{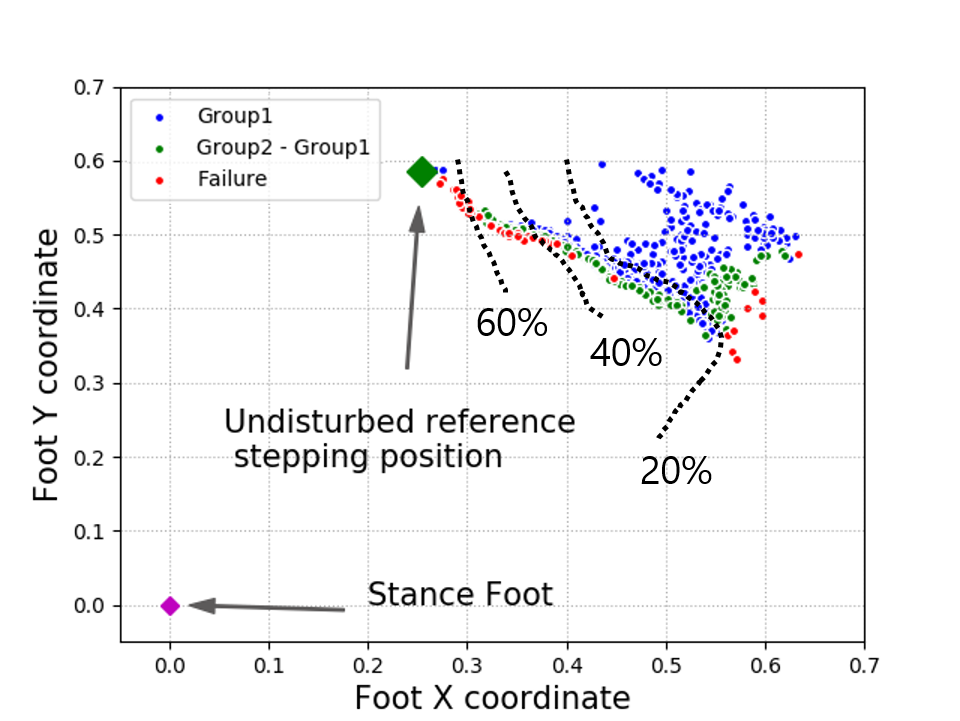}
      \caption{DRL-A* (Normal)}
      \label{fig:foot_placement_push}
    \end{subfigure}
    \begin{subfigure}{.325\textwidth}
      \centering
      \includegraphics[width=\linewidth]{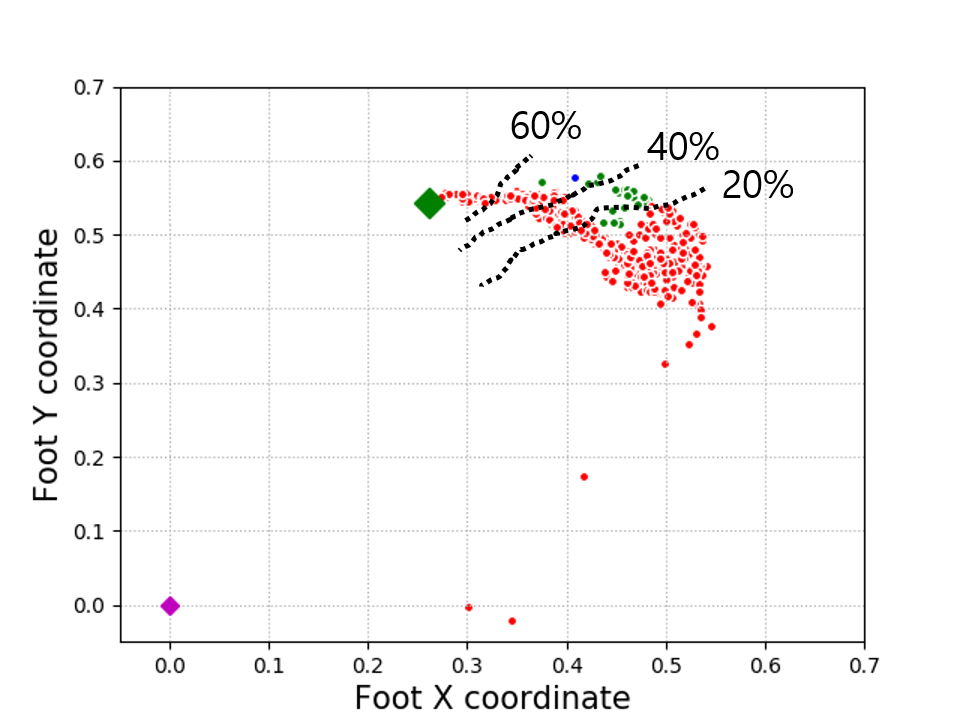}
      \caption{DRL-A (Normal)}
      \label{fig:foot_placement_nopush}
      
    \end{subfigure}
    \vskip\baselineskip
        \begin{subfigure}{.325\textwidth}
      \centering
      \includegraphics[width=\linewidth]{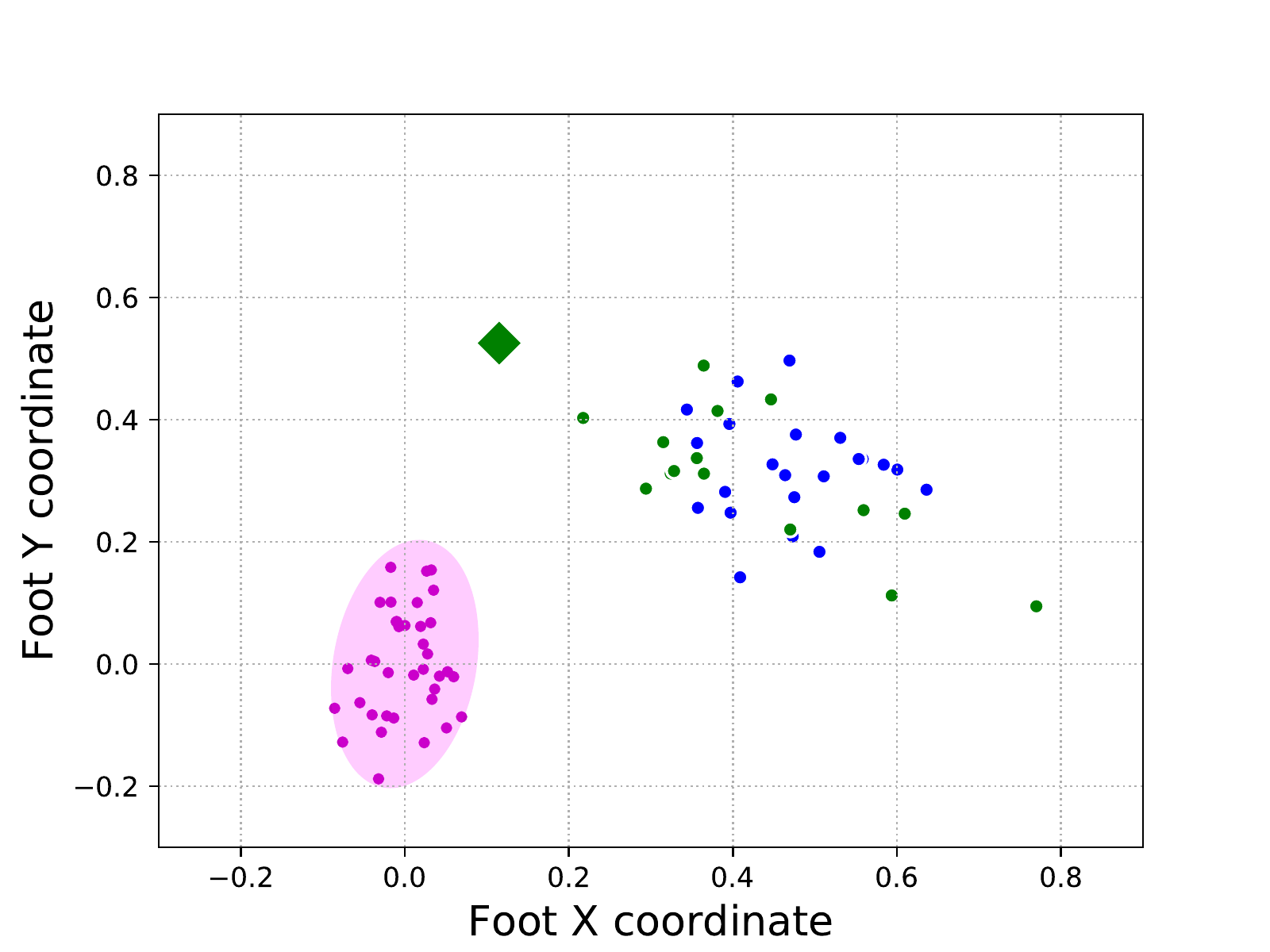}  
      \caption{Human data ($30^{\circ}$-crouch)}
      \label{fig:foot_placement_human_30deg}
    \end{subfigure}
    \begin{subfigure}{.325\textwidth}
      \centering
      \includegraphics[width=\linewidth]{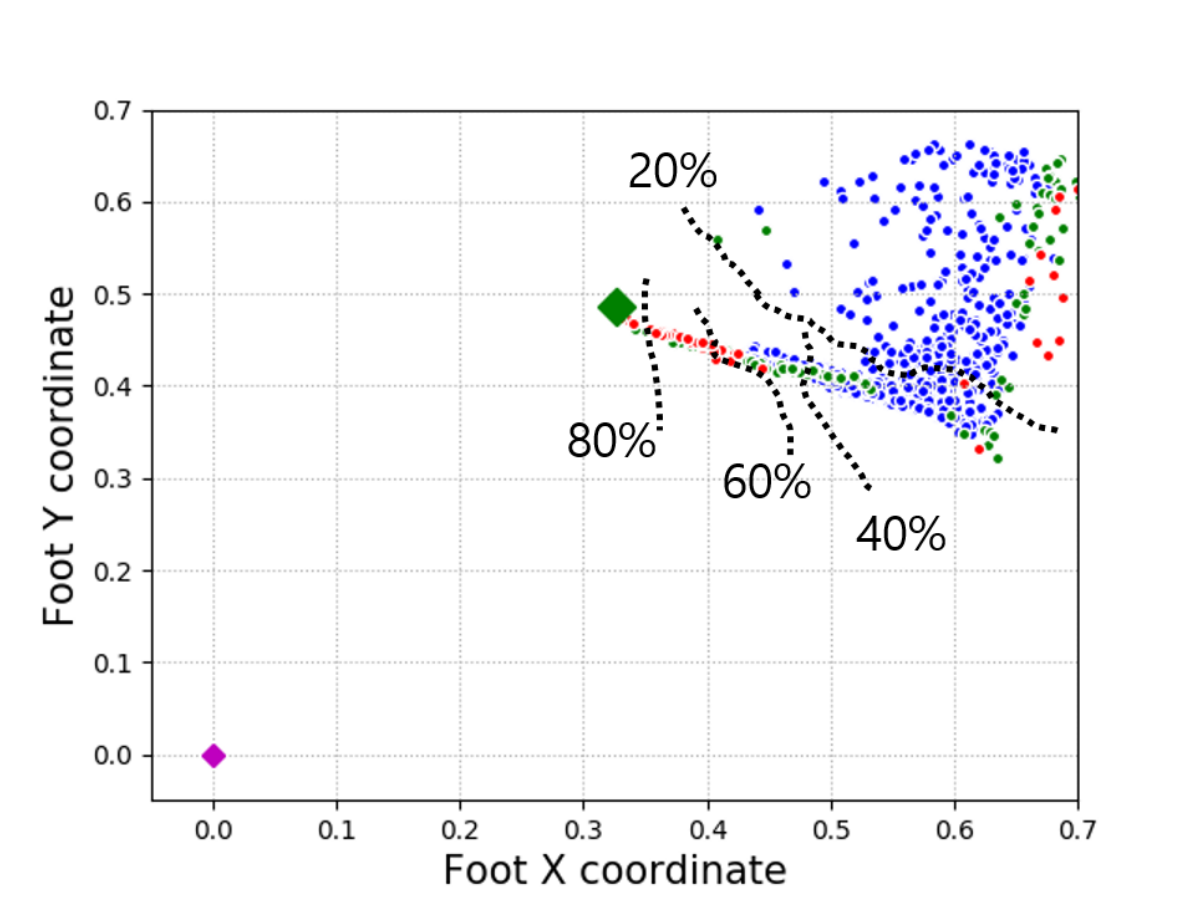}  
      \caption{DRL-A* ($30^{\circ}$-crouch)}
      \label{fig:foot_placement_push_30deg}
    \end{subfigure}
    \begin{subfigure}{.325\textwidth}
      \centering
      \includegraphics[width=\linewidth]{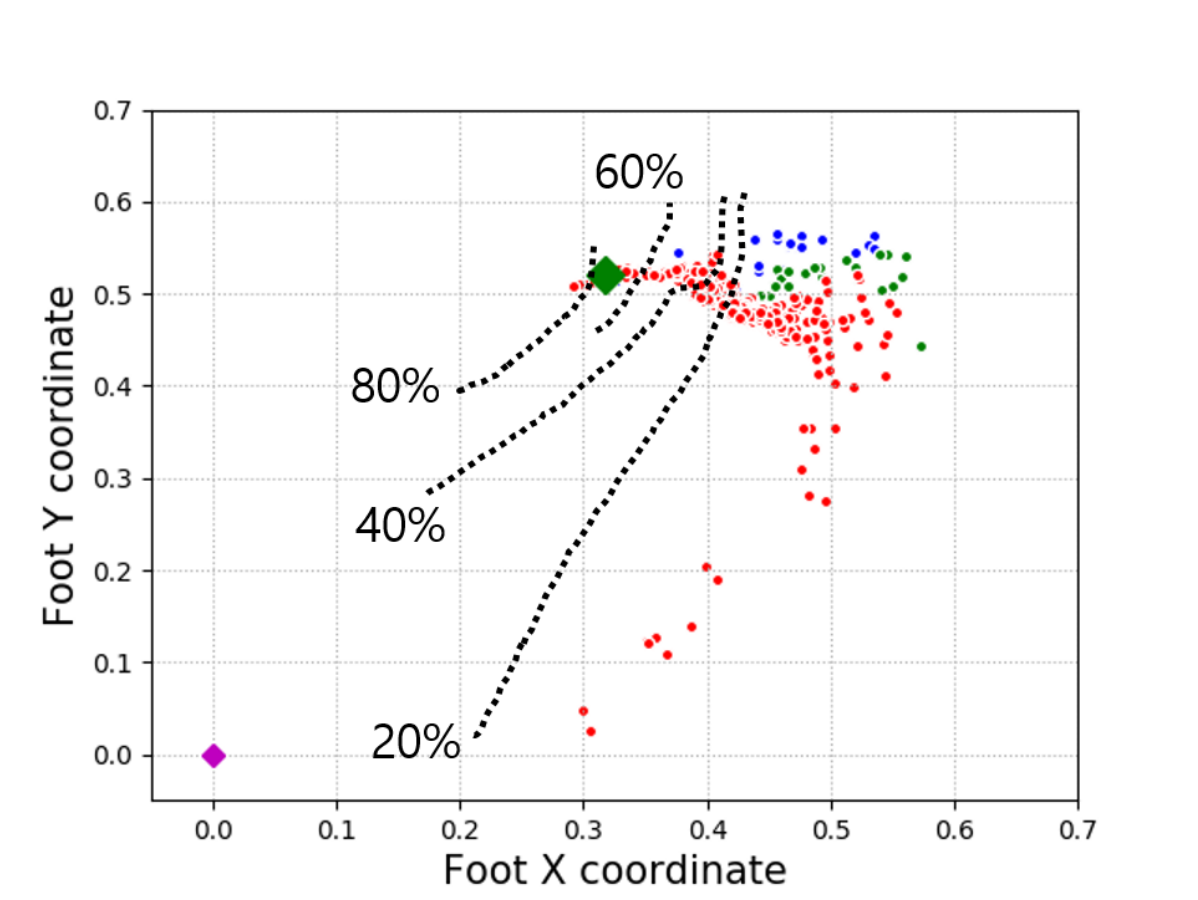}
      \caption{DRL-A ($30^{\circ}$-crouch)}
      \label{fig:foot_placement_nopush_30deg}
    \end{subfigure}
    
    \caption{Foot placement plots. All foot placements are overlaid in the coordinate system. Green diamonds are the footprints to step on in the absence of external disturbances. Pre-push stance footprints are indicated by magenta dots or magenta diamonds. The blue and green dots represent the first post-push footprints of {\em Group 1} and {\em Group 2} - {\em Group 1} in successful episodes, respectively.
    The red dots are those of unsuccessful balance recovery episodes.
 The x-axis and y-axis are lateral and moving directions. The footprints are classified into intervals $(0\%, 20\%, 40\%, 60\%, 80\%, 100\%)$ of push timing indicated by black dotted curves.
 Note that each origin of the coordinates of human data is the mean of stance foot positions.}
    \label{fig:foot_placement_fixed_speed_step_length}
\end{figure*}

It has been previously postulated that balancing strategies based on foot placements play a dominant role in dynamic walking. We hypothesized that deep policies learn to smartly choose post-push footholds to recover balance within a few steps. 
We conducted push-recovery experiments with four deep policies DRL-A, DRL-B, DRL-A*, and DRL-B*. In each trial, the simulated biped was pushed from the left during its left foot in stance and took the next step on the right to recover its balance. Figure~\ref{fig:foot_placement_fixed_speed_step_length} shows that the stability of control policies are characterized by the pattern of post-push foot placements, which are plotted with respect to the reference footprint to step on if there are no disturbances. The push magnitude and timing are randomly sampled, while the walking speed, stride length, and crouch angle are fixed. Push magnitudes are sampled from a normal distribution with a mean of $200N$ and a standard derivation of $35N$. Push timings are sampled from another normal distribution of a mean of $34\%$ and a standard deviation of $21\%$.

We used 158 normal gait data, and 22 30$^\circ$-crouch gait data . 
Note that the human data were collected from many participants under varied conditions and thus necessarily noisy, while the simulation data were collected in a controlled simulation setup. Therefore, the side-by-side direct comparison is meaningless, but the data should be interpreted qualitatively. In the human data (Figure~\ref{fig:foot_placement_human}, \ref{fig:foot_placement_human_30deg}), the disturbed swing foot tends to land behind (in the y-axis) the undisturbed step position. It means that human participants tend to slow down temporarily to cope with lateral disturbances. The success rate of DRL-A is very low ($3.0\%$ for normal walking and $10.1\%$ for 30$^\circ$-crouch walking), and the post-push footholds of successful trials are all at nearly the same y-coordinate as the undisturbed reference step position since DRL-A did not learn how to cope with disturbances. Unlike DRL-A, DRL-A* exploited uncertainty in the learning phase and learned to adjust its step length (shorter in the y-axis) similarly to human balance strategies. 

The foot placement graphs in Figure~\ref{fig:foot_placement_fixed_speed_step_length} also show that post-push step positions are closely correlated with push timing. The push may occur in either a double-stance phase $[0\%,20\%]$ or a swing phase $[20\%,100\%]$. The push at the early swing phase allows the walker to have a sufficient time to move its leg to an appropriate position for balance recovery, while the push at the late swing phase forces the walker to put the swing foot down rapidly with little time for control. Therefore, control policies are more resilient to early-swing pushes than late-swing pushes. Note that maximum detour distances are shorter (can be interpreted as more resilient) with later-swing pushes, which seems contradictory to the proposition based on the success rate. As we discussed before, there is no ultimate criterion for resilience and gait stability. The notion of stability can be characterized richly by a mixture of measures.
For example, Figure~\ref{fig:foot_placement_fixed_speed_step_length} shows that the post-push step positions tend to be farther from the undisturbed reference step position as the timing of the push is earlier (the 20\% curve is farther than the 60\% curve from the reference position), but the success rate for early pushes are higher than late pushes (the ratio of blue and green dots in the <20\% push timing region is higher than that of the 40\%--60\% region).

As shown in Figure~\ref{fig:foot_placement_fixed_speed_step_length}, the successful post-push steps of normal gaits are mostly placed less in the y-axis than the reference stepping  position, as in the human data. 30$^\circ$-crouch gaits are not necessarily the case, but the y values of more successful post-push steps are smaller than that of the reference position. This means that perturbation during the learning process allows learning policies that are more similar to the actual human balancing mechanism. The significant fatigue experienced by the human participants during crouch walking is not modeled in our simulation model, which might result in the slightly different tendency of post-push steps  for 30$^\circ$-crouch gaits.



\begin{figure}
\begin{subfigure}{.235\textwidth}
  \centering
  \includegraphics[width=\linewidth]{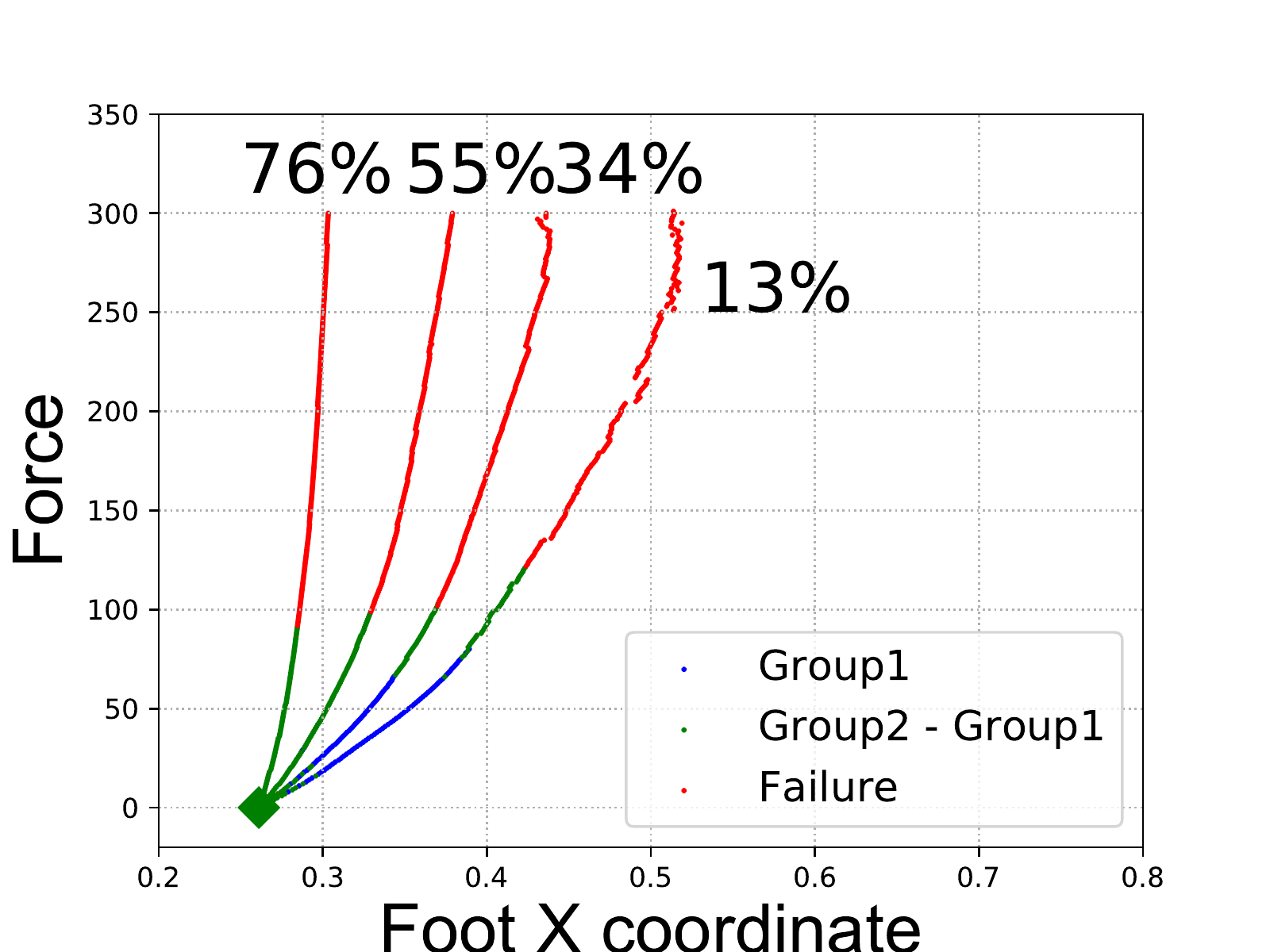}  
  \caption{DRL-A (Pushing direction)}
  \label{fig:foot_placement_nopush_x}
\end{subfigure}
\begin{subfigure}{.235\textwidth}
  \centering
  \includegraphics[width=\linewidth]{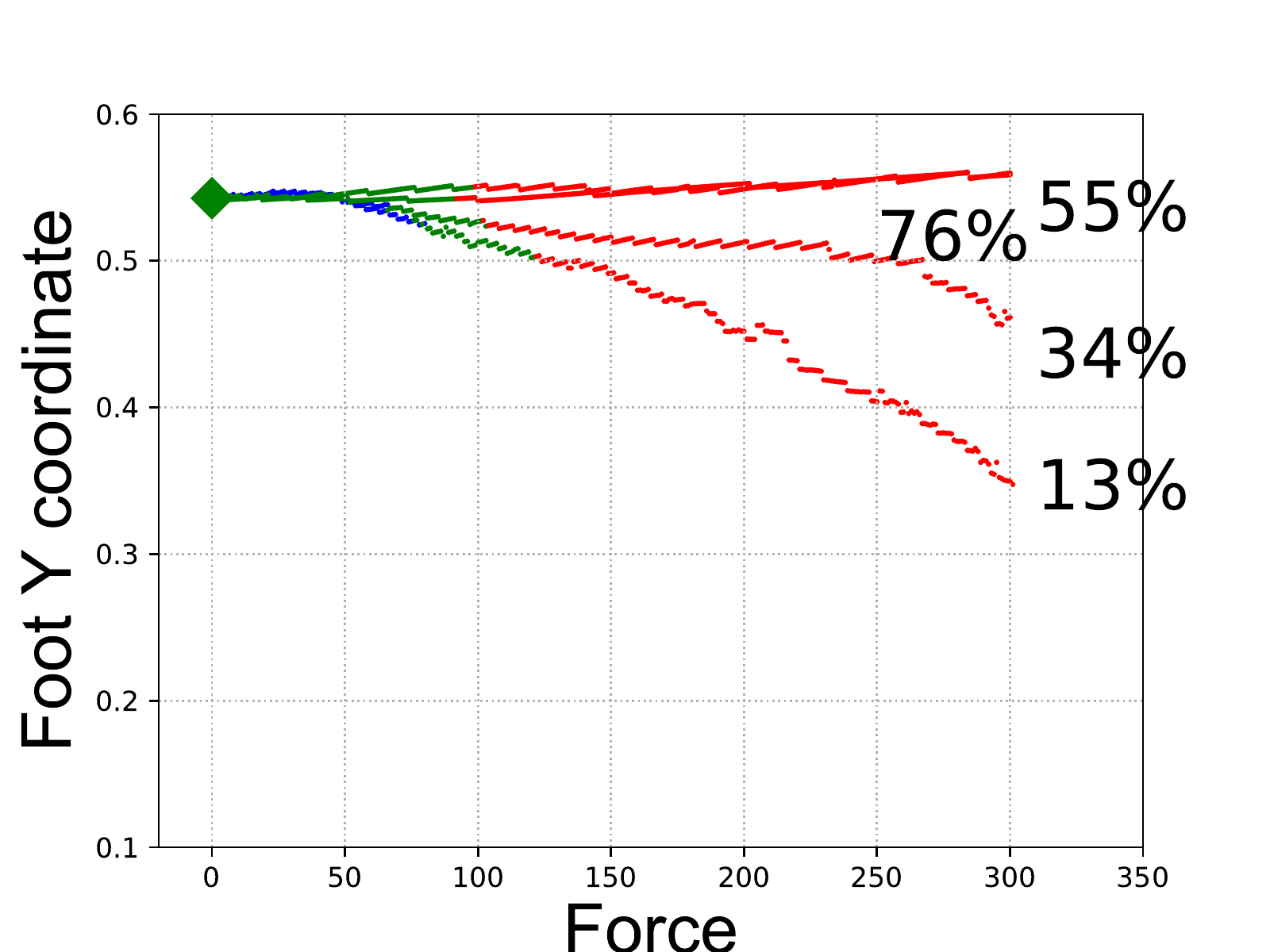}  
  \caption{DRL-A (Walking direction)}
  \label{fig:foot_placement_nopush_y}
\end{subfigure}
\vskip\baselineskip
\begin{subfigure}{.235\textwidth}
  \centering
  \includegraphics[width=\linewidth]{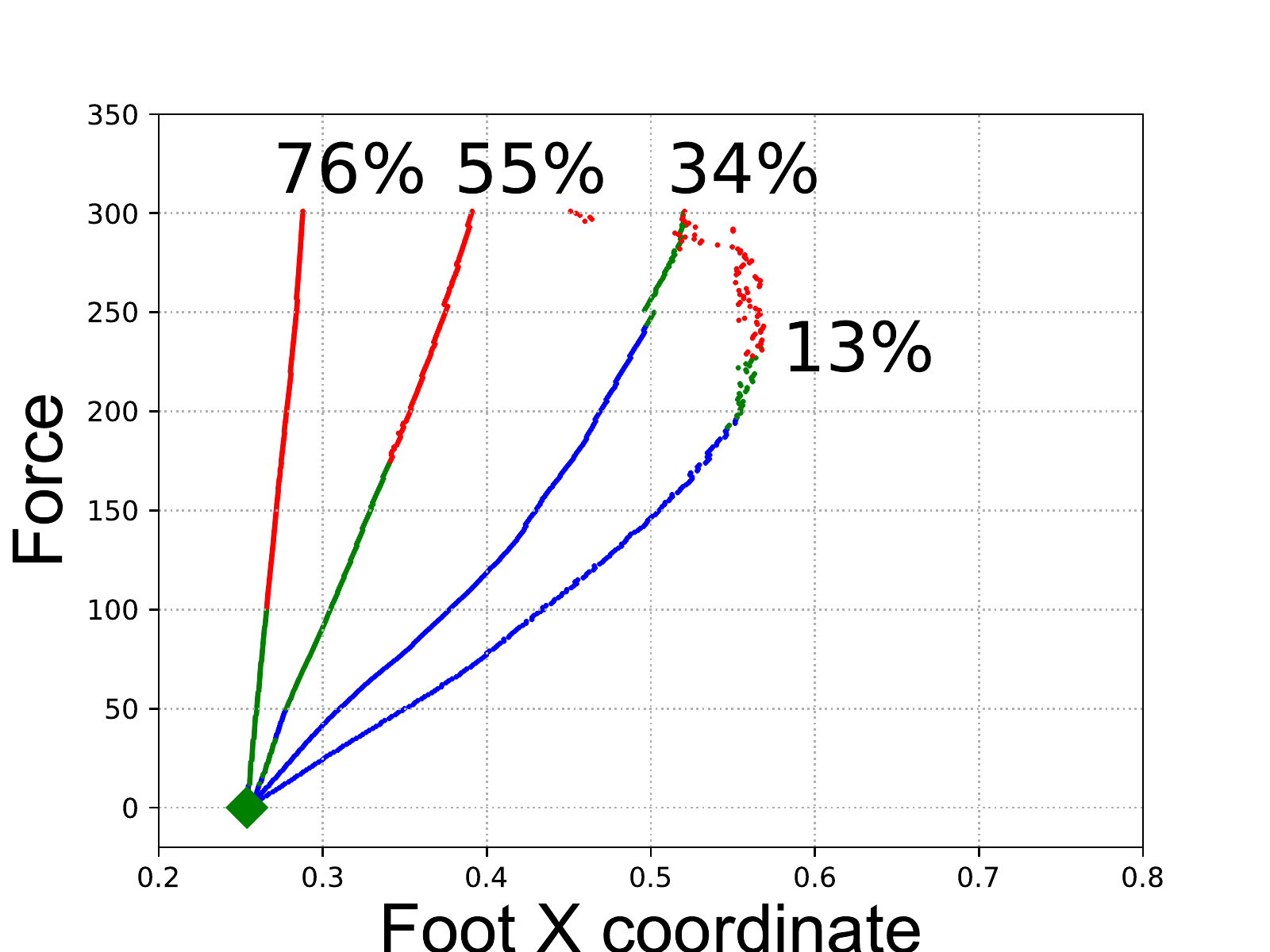}  
  \caption{DRL-A* (Pushing direction)}
  \label{fig:foot_placement_push_x}
\end{subfigure}
\begin{subfigure}{.235\textwidth}
  \centering
  \includegraphics[width=\linewidth]{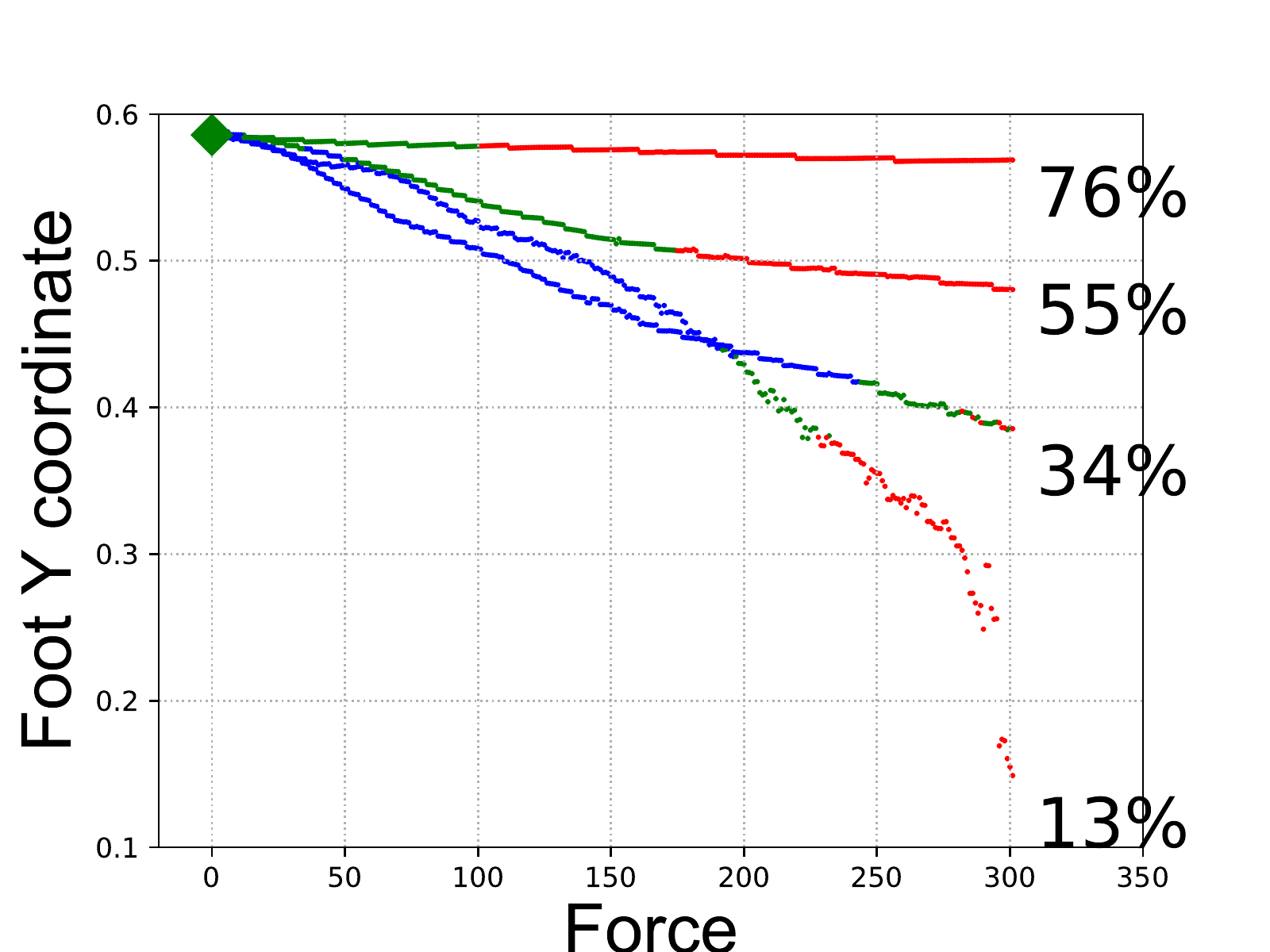}  
  \caption{DRL-A* (Walking direction)}
  \label{fig:foot_placement_push_y}
\end{subfigure}
\caption{Foot placement plots at four push timings (13\%, 34\%, 55\%, 67\%). Green diamond indicates undisturbed reference step position. }
\label{fig:foot_placement}
\end{figure}

The plots in Figure~\ref{fig:foot_placement} depict the correlation between post-push foot placements and push magnitude/timing. DRL-A* learned its balance strategies based on foot placements, which are strongly correlated with how strong the disturbance is and when it occurs. DRL-A* is more robust than DRL-A at all push magnitudes and timings. DRL-A* adjusts its post-push foothold not only in the push (lateral) direction but also in the moving direction to better respond to the pushes.
We also observed the clear correlation of the number of detour steps with push magnitude. As push magnitude increases, the character needs one step ({\em Group 1} marked in blue) to three steps ({\em Group 2} - {\em Group 1} marked in green) to recover balance after the push.  

\subsection{Application: Interactive Control}

In an interactive control problem, it is critical to process the user input on-the-fly. A sudden change of the reference motion during the simulation can act as an impulse in the system. 
Since the gait-conditioned policies learn from diverse walking motions under disturbances, a simulated character controlled by these policies is robust to the change of input motions, and thus can transit from a gait pattern to another smoothly without any additional technique such as motion blending.

\begin{figure}
    \centering
    \includegraphics[width=\columnwidth]{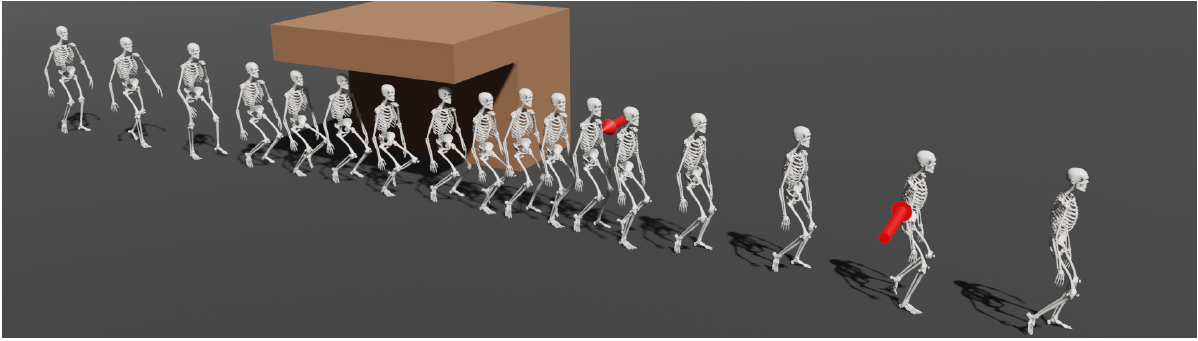}
    \caption{Interactive control. By using a single gait-conditioned policy, we can simulate a sequence of the biped walking motions that span a wide range of walking speeds, stride lengths, and crouch angles as well as the external pushes.}
    \label{fig:interactive_control}
\end{figure}

We demonstrate an interactive control to show the robustness of our gait-conditioned policies (Figure~\ref{fig:interactive_control}). In this example, we compare the four deep control policies: DRL-A*, DRL-B*, DRL-A, and DRL-B. Each character is controlled by a single control policy, and the reference motion is changed whenever the user input consisting of walking speed, stride length, and crouch angle is given. Despite diverse types of sabotaging with the obstacle, push forces, and repeated changes of input motion, the character controlled by DRL-A* survives alone to the last. Please refer the
supplemental video at (05:24). 

\section{Discussion}

Biped locomotion control has been a long-standing problem in computer graphics. Tremendous efforts have been put into addressing this problem for decades and the computer graphics community has finally reached to the point where simulated controllers are worth comparing with the performance and capability of humans in terms of robustness, energy-efficiency, agility, diversity, and flexibility. This study would be a stepping stone towards rigorous evaluation of DRL policies for biped locomotion.

Although our experiments demonstrated the push-recovery stability of DRL policies in varied conditions, there are still numerous gait factors, body/environment conditions, types of disturbances, technical issues, and limitations to be addressed in future studies. Currently, the effects of lateral disturbances are evaluated in our study. We can think of many other conditions including disturbances at arbitrary locations and directions, and responses to slipping, tripping, slopes, and uneven terrain~\cite{peng_deeploco:_2017, Wang:2010:SIGGRAPH}. 

Physiological factors such as effort and fatigue are not considered yet. The dynamics model of our bipeds does not have explicit torque limits. Without torque limits, the biped can withstand arbitrarily strong disturbances with unrealistically fast and strong responses. Fortunately, such an unnatural situation does not occur in our experiments probably because joint torques are implicitly limited in the reference-tracking framework. It is strenuous for human participants to walk in $60^\circ$-crouch. $60^\circ$-crouch walking is less stable than $30^\circ$-crouch walking probably due to fatigue, which is not implemented in our DRL algorithms. The simulated bipeds never get tired in strenuous tasks. Incorporating effort and fatigue into a dynamical system requires the concept of energy/torque minimization, which plays a central role in trajectory optimization algorithms~\cite{al2012trajectory, lee2014locomotion}. However, it is still unclear how the concept can be implemented in the DRL framework. Learning energy-efficient, compliant policies would be an interesting direction for future studies.

In our work, we parameterize the reference motion clip to simulate various walking motions with crouch angles, walking speed, and stride length.
Then the motion is warped according to selected gait parameters and then used to train a control policy.
The problem is that, unlike the unwarped reference motion that reflects the dynamics of character, warped motions might not be compatible with the model’s dynamics.
However, our gait-conditioned policy can mimic various motions warped by gait parameters. 
This is possible because there is a common control method required to mimic walking motions, and this common control method can be learned through DRL.
The control method can be learned more easily from the motions close to the model's dynamics and serves as a guide for the largely warped motions far from the dynamics.




Lessons can be drawn from our experiments for the evaluation and implementation of DRL algorithms. First, gait and stability are strongly correlated. Crouching and stomping gaits that might look unnatural or impaired would be more robust than typical, normal gaits. Therefore, evaluating the robustness and stability of the controller entails gait normalization for a fair comparison. Secondly, including uncertainty and disturbances in the learning phase is imperative for learning robust control policies that mimic human balance strategies. Thirdly, gait-conditioned policies are very useful in terms of computation time and memory usage. The use of adaptive sampling is strongly encouraged to improve the robustness of control policies uniformly across a range of parametric domain.

There are many promising applications we can think of. Interactive graphics applications, such as video games, often show animated controllable characters that hit, react, and fall. Understanding their resilience against disturbances would be useful for better simulation of their interactions. We can also think of an exoskeleton-type walking assist device that generates assist force/torque at the joint of the wearer. Many devices are designed to help the elderly and the handicapped who are exposed to the risk of falls. Walking assist devices equipped with the ability to prevent falls would be highly desirable~\cite{constantinescu2007assistive, low2011robot}. We believe that our analysis in this study will provide a solid basis for designing such devices.
\bibliographystyle{eg-alpha-doi}  
\bibliography{EGauthorGuidelines-sca2020-sub}        


\end{document}